\documentclass[prx,superscriptaddress,aps, bibliography, longbibliography, twocolumn, reprint, showpacs, footnoteinbib]{revtex4-1}

\usepackage{enumerate}
\usepackage{braket}  
\usepackage{graphicx}
\usepackage{amssymb} 
\usepackage{amsfonts}
\usepackage{amsmath}
\usepackage{dsfont}
\usepackage{natbib}
\usepackage{dcolumn} 
\usepackage{bm}  
\usepackage[mathscr]{eucal}
\usepackage[dvipsnames]{xcolor}
\usepackage[colorlinks, linkcolor=OrangeRed,citecolor=RoyalBlue,urlcolor=NavyBlue]{hyperref}
\usepackage[all]{hypcap} 
\usepackage{mathtools}
\usepackage{wasysym}

\newcommand{\mc}{\mathcal}

\newcommand{\ep}{\varepsilon}

\newcommand{\be}{\begin{equation}}
\newcommand{\ee}{\end{equation}}
\def\ba{\begin{aligned}}
\def\ea{\end{aligned}}
\newcommand{\bea}{\begin{eqnarray}}
\newcommand{\eea}{\end{eqnarray}}
\def\bes{\begin{subequations}}
\def\ees{\end{subequations}}
\def\bal{\begin{align}}
\def\eal{\end{align}}

\newcommand{\la}{\left\langle}
\newcommand{\ra}{\right\rangle}
\newcommand{\lv}{\left|}
\newcommand{\rv}{\right|}
\newcommand{\lb}{\left[}
\newcommand{\rb}{\right]}
\newcommand{\lp}{\left(}
\newcommand{\rp}{\right)}

\renewcommand{\hat}[1]{{\widehat #1}}

\newcommand{\tr}{{\rm Tr}}

\usepackage{soul}
\usepackage{ulem} 

\graphicspath{{images/}}

\begin{document}

\title{Multifractality meets entanglement: relation for non-ergodic extended states}
\author{Giuseppe De Tomasi}
\affiliation{T.C.M. Group, Cavendish Laboratory, JJ Thomson Avenue, Cambridge CB3 0HE, United Kingdom}
\affiliation{Department of Physics, Technische Universit\"at M\"unchen, 85747 Garching, Germany}
\author{Ivan M. Khaymovich}
\affiliation{Max-Planck-Institut f\"ur Physik komplexer Systeme, N\"othnitzer Stra{\ss}e 38, 01187-Dresden, Germany}
\begin{abstract}
In this work we establish a relation between entanglement entropy and fractal dimension $D$ of generic many-body wave functions, by generalizing the
result of Don N. Page [Phys. Rev. Lett. 71, 1291] to the case of {\it sparse} random pure states (S-RPS).
These S-RPS living in a Hilbert space of size $N$ are defined as normalized vectors with only $N^D$ ($0 \le D \le 1$) random non-zero elements.
For $D=1$ these states used by Page represent ergodic states at infinite temperature.
However, for $0<D<1$ the S-RPS are non-ergodic and fractal as they are confined in a vanishing ratio $N^D/N$ of the full Hilbert space.
Both analytically and numerically, we show that the mean entanglement entropy ${\mathcal{S}_1}(A)$ of a subsystem $A$, with Hilbert space dimension $N_A$, scales as
$\overline{\mathcal{S}_1}(A)\sim D\ln N$ for small fractal dimensions $D$, $N^D< N_A$.
Remarkably, $\overline{\mathcal{S}_1}(A)$ saturates at
its thermal (Page) value at infinite temperature, $\overline{\mathcal{S}_1}(A)\sim \ln N_A$
at larger $D$.
Consequently, we provide an example when the entanglement entropy takes an ergodic value even though the wave function is highly non-ergodic.
Finally, we generalize our results to Renyi entropies $\mathcal{S}_q(A)$ with $q>1$ and to genuine multifractal states and also show that their fluctuations have ergodic behavior in narrower vicinity of the ergodic state, $D=1$.

\end{abstract}
\maketitle
\textit{Introduction}--
The success of classical statistical physics is based on the concept of ergodicity, which allows the description of complex systems by the knowledge of only few thermodynamic parameters~\cite{landau2013statistical, boltzmann1896vorlesungen}.
In quantum realm the paradigm of ergodicity is much less understood and its characterization is now an active research front.

The most accredited theory, which gives an attempt to explain equilibration in closed quantum systems, relies on the  eigenstate thermalization hypothesis (ETH)~\cite{Deutsch_1991, Sredni94, Srednicki_1996, Rigol2008}. ETH assets that the system thermalizes locally at the level of single eigenstates and has been tested numerically in a wide variety of generic interacting systems~\cite{Review_Rigol_16, Rigol2008}.

It is now well established that entanglement plays a fundamental role on the thermalization process~\cite{Nandkishore_review_15, Abanin_review_19, ALET2018498, Review_Rigol_16, Amico_08}. Thermal states are locally highly entangled with the rest of the system, which acts as a bath. Consequently, the measurement of entanglement entropy (EE)  has been found to be a resounding resource to probe ergodic/thermal phases, both theoretically~\cite{Luitz15, Kjall_2014, Gara_2016, Abanin_power_law_spectrum_2016, GDT_QMI_2017, Critical_Vedika_2017} and recently also experimentally~\cite{Islam2015, Kaufman794, MBL_exp_2019, Lukin256}.
For instance, infinite temperature ergodic states are believed to behave like random vectors~\cite{Deutsch_1991,Review_Rigol_16}  and their EE reaches a precise value often referred as Page value~\cite{page1993}.

On the other hand, ergodicity is deeply connected to the notion of chaos~\cite{Haake_chaos,Review_Rigol_16}, which implies also an equipartition of the many-body wave function over the available  many-body Fock states, usually quantified by multifractal analysis, e.g., by scaling of the inverse participation ratio (IPR)~\cite{Evers2008Review}. In this case, infinite temperature ergodic states span homogeneously the entire Hilbert space~\cite{WE-footnote}. The latter states should be distinguished from the so-called non-ergodic extended (NEE) states.
These NEE states live on a fractal in the Fock space, which is a vanishing portion of the total Hilbert space. Recently, the NEE have been invoked to understand new phases of matter like \textit{bad metals}~\cite{DeLuca14,Kravtsov_NJP2015,Alt16,pino2017mulitfractal,RRGAnnals, Bera_2018, GDT_2019_RRG, GDT_return_2019,LN-RP2020}, which are neither insulators nor conventional diffusive metals and also are found in chaotic many-body quantum system like in the Sachdev-Ye-Kitaev model~\cite{micklitz2019non,Kamenev-talk-1,Kamenev-talk-2}.
%

Very recently, the two aforementioned probes, EE and IPR, have been used to describe thermal phases (specially at infinite temperature), and to detect ergodic-breaking quantum phase transitions~\cite{serbyn2013universal,Luitz15,luitz2019multifractality, Mace_2019, Monthus_2016}.
Nevertheless, the relations between these two probes has not been studied extensively so far~\cite{EE_IPR_footnote}.
Thus, the natural question arises: to what extend do these probes lead to the same description?

In this work, we build up a bridge between ergodic properties extracted from EE and the ones from multifractal analysis. With this aim, we generalize the seminal work of Page~\cite{page1993}, computing EE and its fluctuations for NEE states.
Remarkably, we show, both analytically and numerically, that a subsystem EE 
can still be ergodic (Page value),
even though the states are highly non-ergodic.
Consequently, the mean value of EE might be not enough to state ergodicity, though EE reaches the Page value.

\textit{General definitions}--
The Renyi entropy, $\mathcal{S}_q(A)$, of a subsystem $A$ with Hilbert space dimensions $N_A = N^p$, $p\leq 1/2$, is defined as:
\be
\label{eq:S_q}
\mathcal{S}_q(A) = \frac{\ln \Sigma_q}{1-q} \ , \text{ with } \Sigma_q = \tr_A[\rho_A^q]=\sum_{M=1}^{N_A} \lambda_M^q, \
\ee
where  $\rho_A = \tr_B[\rho ]$ is the reduced density matrix of the subsystem $A
$, after tracing out the subsystem $B=A^{c}$ and $\{\lambda_M\}$ are Schmidt eigenvalues of $\rho_A$.
The von Neumann EE, $\mathcal{S}_1(A) = \lim_{q\rightarrow 1} \mathcal{S}_1(A)$ equals to  $-\tr_A[ \rho_A\ln{\rho_A}]$.
For a pure state $\rho = \lv \psi \ra \la \psi \rv$, 
\be\label{eq:rho_A}
\rho^A_{M,M'} = \sum_{m=1}^{N_B} \psi_{M,m}\psi_{M',m}^* \ ,
\ee
where $\psi_{M,m}$ are the wave function coefficients $\lv \psi \ra = \sum_{M=0}^{N_A-1}\sum_{m=0}^{N_B-1} \psi_{M,m} \lv M \ra_A\otimes\lv m \ra_B$
in the computational basis $\lv M \ra_A$, $1\leq M\leq{N_A}
$, and $\lv m \ra_B$, $1\leq m\leq {N_B}$, of the two subsystems $A$ and $B$, respectively.

For fully random states, $D=1$, the mean von Neumann EE is given by the Page value~\cite{page1993}
\be\label{eq:Page}
\overline{\mathcal{S}_{{\rm Page}}}(A)  = \ln N_A - \frac{N_A}{2 N_B},
\ee
and its fluctuations decays to zero,
$\overline{\delta \mathcal{S}_{{\rm Page}}}(A) =  (\overline{\mathcal{S}^2}(A)- \overline{\mathcal{S}}^2(A))^{1/2} \sim N_B^{-1}$~\cite{Majumdar2010RPS,Majumdar2011RPS,Vivo2016RPS}.
The overline indicates the random vector average.

Moreover, the ergodic properties of the wave function $\{\psi_{\bold{n} = (M,m)}\}$ can be characterized in terms of multifractal analysis~\cite{Evers2008Review}
via the fractal dimensions $D_q$, ${q\ge 0}
$, defined through the scaling of the inverse participation ratios ${\rm IPR}_q$ with $N$,
\be
\label{eq:D_q}
D_q \ln {N} = \frac{\ln {\rm IPR}_q }{1-q}  \ , \text{ with } {\rm IPR}_q=\sum_{\bold{n}}|\psi_\bold{n}|^{2q},
\ee
giving in the limit $q\rightarrow 1$, $D_1 \ln {N} = -\sum_{\bold{n}} |\psi_\bold{n}|^{2}\ln |\psi_\bold{n}|^{2}
$.

The exponent $D_1$ provides important information on the dimension of the support set of the wave function in the Fock space, which scales as $\sim N^{D_1}$~\cite{KravtsovSupportSet}.
Ergodic states are characterized by $D_q=1$, meaning that the state is homogeneously spread over the entire Hilbert space~\cite{WE-footnote}. Instead, NEE states are usually multifractal with $D_q<1
$ and their support set is a vanishing ratio of the full Hilbert space $\sim N^{D_1}/N$.

\begin{figure}[h!]
 \includegraphics[width=1.\columnwidth]{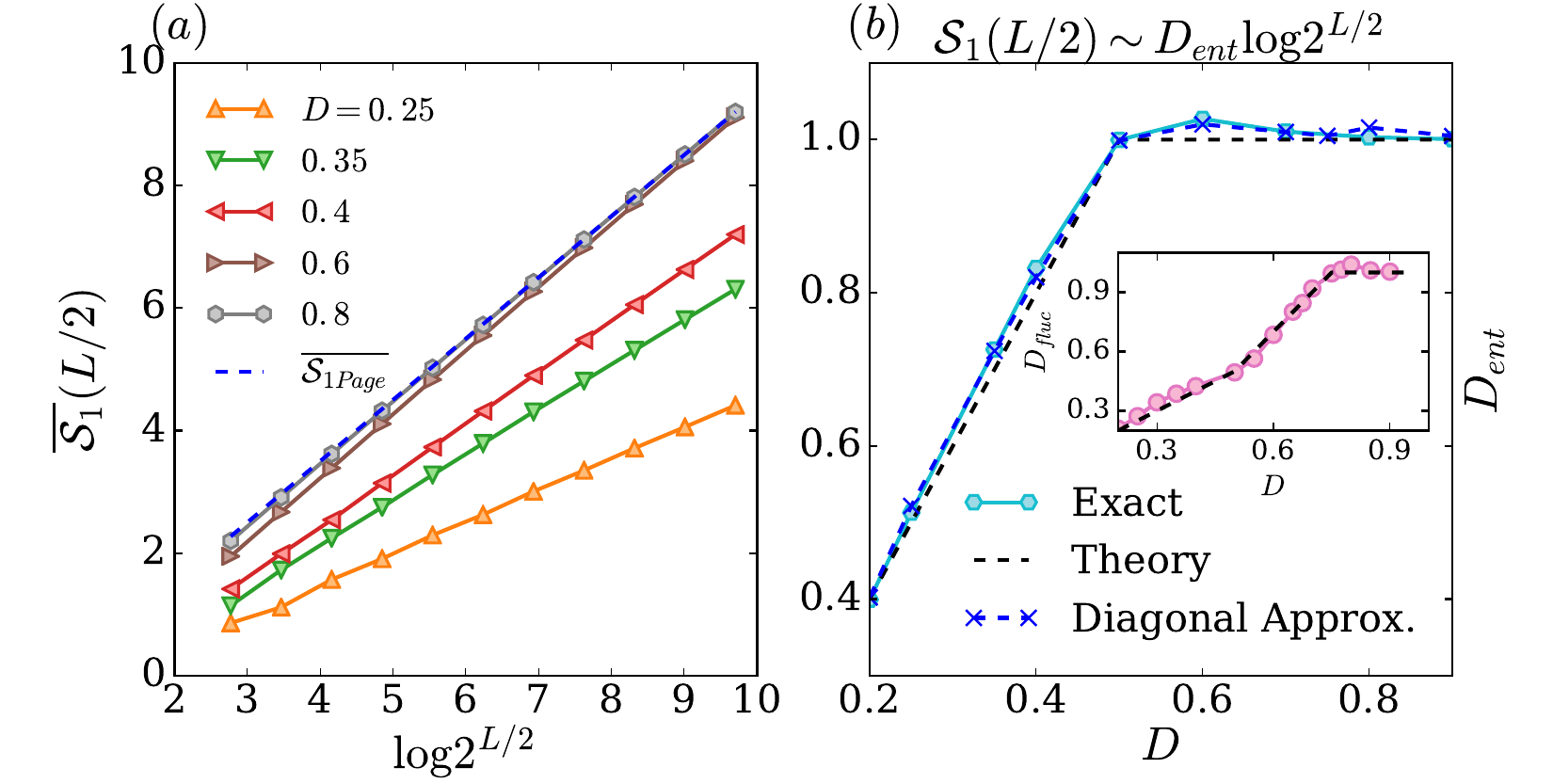}
 \caption{{\bf von Neumann EE scaling versus fractal dimension $D$ for S-RPS}.
 (a)~$\overline{\mathcal{S}_{1}}(L/2)$ of half-system, $N_A=2^{L/2}$, versus $L$ for different $D$; dashed line shows the Page value, Eq.~\eqref{eq:Page}.
 (b)~Slope $D_{ent}$ of $\overline{\mathcal{S}_{1}}(L/2) \sim D_{ent} L/2\ln{2}$ versus $D$ (Exact) and of $-\sum_i \rho_{M,M} \ln{\rho_{M,M}} \sim D_{ent} L/2\ln{2}$ (Diagonal Approx.); black line represents the theoretical prediction in Eq.~\eqref{eq:S_1_analytic}.
 (inset)~Slope $D_{fluc}$ of the standard deviation $\overline{\delta \mathcal{S}}(L/2) \sim D_{fluc} \ln{2^{L/2}}$ versus $D$.
 Black dashed line shows analytical prediction, Eq.~\eqref{eq:S_fluct}, for $p=1/2$.
 }
\label{fig:fig2}
 \end{figure}


\begin{figure*}[t!]
 \includegraphics[width=1.\textwidth]{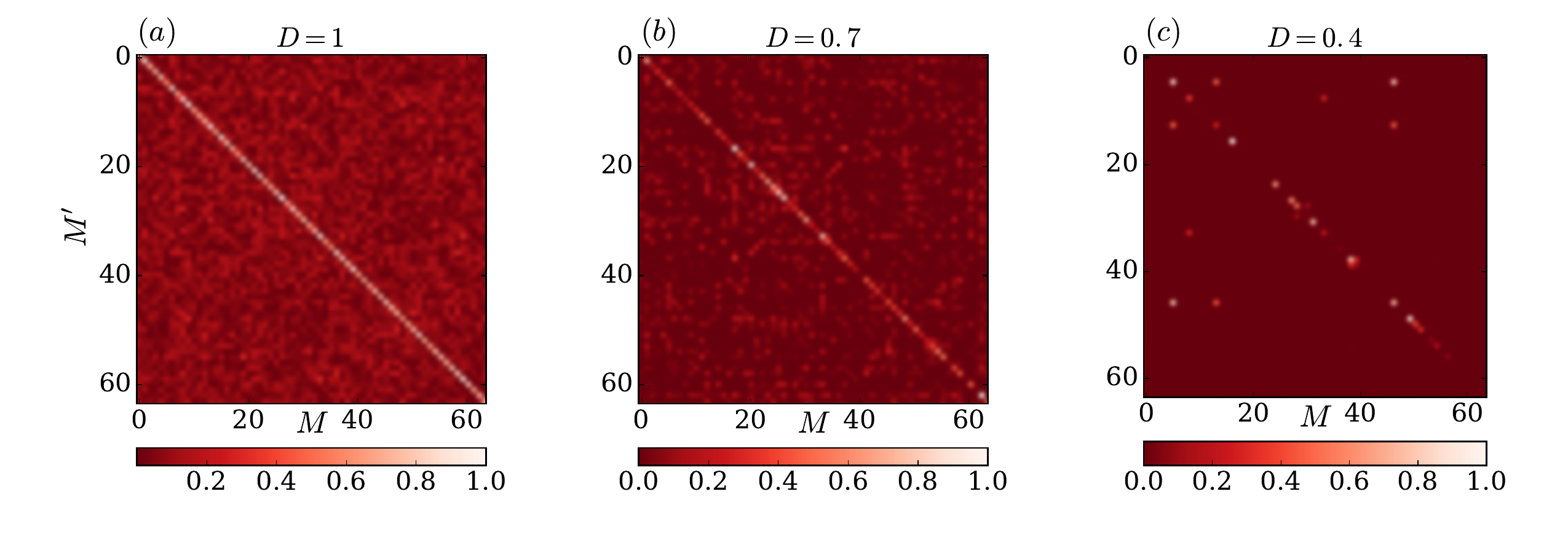}
 \caption{{\bf Structure of half-system reduced density matrix} $|\rho_A^{M,M'}|/\max_{M,M'}|\rho_A^{M,M'}|$ for S-RPS with
 (a)~$D=1$, (b)~$D=0.7$, and (c)~$D=0.4$ for $N_A=2^{L/2}$ and $L=12$.
 In all panels $\rho_A$ is mostly represented by the diagonal elements with almost uniform distribution $\rho_A^{M,M} \sim N_A^{-1}$ for $D>1/2$ (a, b)
 and bimodal distribution otherwise~(c).
 The latter case is given by $\sim 2^{DL}$ non-zero nearly uniform elements normalized as $\rho_A^{M,M} \sim 2^{-DL}$ with the rest being negligibly small.
 The corresponding EE saturates at the ergodic Page value $\overline{\mathcal{S}_1}(A)=\overline{\mathcal{S}_{\rm Page}}(A)$ for $D>1/2$, while being dominated by $2^{DL}$ non-zero elements for $D<1/2$ leading to
$\overline{\mathcal{S}_1}(A)\simeq -\sum_M \rho_A^{M,M} \ln{\rho_A^{M,M}} \sim D L\ln{2}$.
}
 \label{fig:fig1}
 \end{figure*}

In this work, we consider entanglement properties of NEE states by introducing the {\it sparse} random pure states (S-RPS).
The S-RPS are normalized random vectors $\{\psi_\bold{n}\}$ with $N^D$ non-zero elements, that are randomly distributed over the Hilbert space. 
Similarly to RPS~\cite{page1993} ($D=1$) all non-zero coefficients are Gaussian distributed with the normalization-controlled width.
The S-RPS are described by only one fractal dimension $D_q=D<1
$, ${\rm IPR}_q \sim N^{D(1-q)}$, $q>0$~\cite{MF-footnote}. Thus, the S-RPS are homogeneously spread, but only in a vanishing ratio of the total Hilbert space.
\begin{figure}[h!]
 \includegraphics[width=1.\columnwidth]{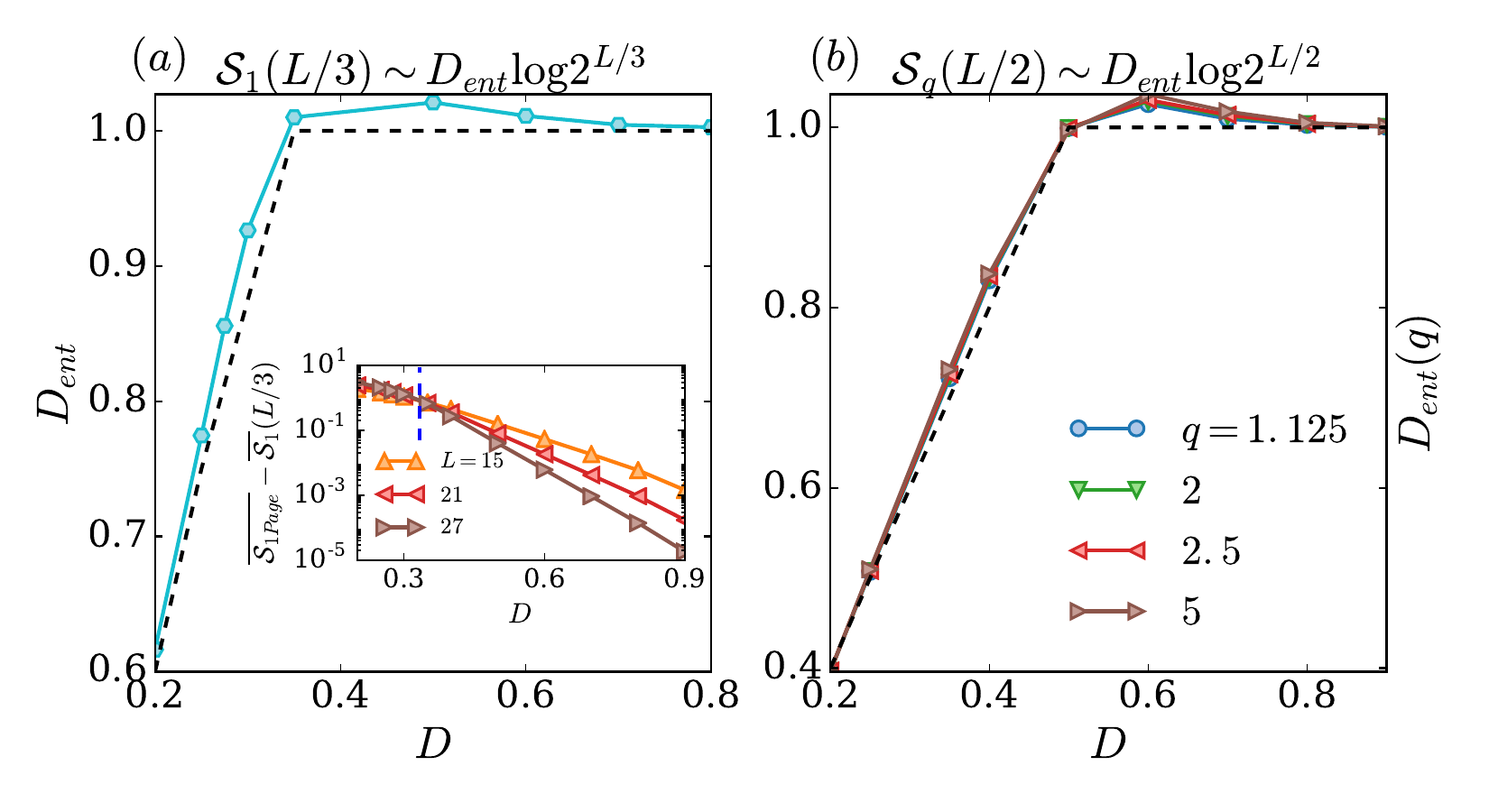}
 \caption{{\bf Effect of partition size and scaling of Renyi EE}.
 (a)~Slope $D_{ent}$ of the mean EE versus fractal dimension $D$  with $N_A=2^{L/3}$.
 (inset) $\overline{\mathcal{S}_{\rm Page}}(L/3) -\overline{\mathcal{S}_{1}}(L/3)$ versus $D$ showing the corrections to the Page value exponentially suppressed with $L$ for $D>p=1/3$ (vertical dashed line).
 (b)~Slope $D_{ent}(q)$ of the mean Renyi EE of a half-system for different $q$.
 The dashed blacks lines represent theoretical predictions in Eq.~\eqref{eq:S_1_analytic}.
 }
\label{fig:fig21}
 \end{figure}

\textit{Results}--- We start to outline our results, by  computing numerically the mean EE for S-RPS with fractal dimension $0<D<1$ in a Hilbert space of dimension $N=2^{L}$~\footnote{For more system examples, including ones with genuine multifractality, please see Appendix~\ref{App_Sec:num_egs}}. In this case, the S-RPS could be thought as eigenstates in the middle of the spectrum~\footnote{Thus, local observables thermalize to their infinite-temperature values.} of some strongly interacting spin-$\tfrac12$ chain with $L
$ sites.

First, we consider limiting cases:
for $D=1$, $\overline{\mathcal{S}_1}(A)$ is given by the Page value, Eq.~\eqref{eq:Page},
$\sim \ln{N_A}$, as the system is ergodic,
while for $D=0$, the wave function is localized in the Fock-space and EE shows area-law $\overline{\mathcal{S}_1}(A)\sim \mathcal{O}(1)$. For $0<D<1$, one may expect the natural interpolation  $\overline{\mathcal{S}_1}(A)\sim D\ln{N_A}$, as S-RPS are random states in a sub-Hilbert space of dimension $N^D$.
However, as we will show, this intuitive picture is misleading.

%

Figure~\ref{fig:fig2} presents the mean value of the half-partition EE, $\overline{\mathcal{S}_1}(L/2)$, $N_A=2 ^{L/2}$.
$\overline{\mathcal{S}_1}(L/2)$ follows a volume law
$\mathcal{S}_1(L/2) \sim D_{ent} \ln{2^{L/2}}$ for any $D>0$ and the slope $D_{ent}$ grows with increasing $D$.
However, the curves approach the Page value $\overline{\mathcal{S}_{1 Page}}(L/2) = L/2\ln{2}-1/2$ (dashed line in Fig.~\ref{fig:fig2}~(a)), i.e., $D_{ent} = 1$ for $D>1/2
$.
Instead, for $D<1/2$,  $\overline{\mathcal{S}_1}(L/2)$ grows slower than $\overline{\mathcal{S}}_{1 Page}(L/2)$ and we found $D_{ent}=2D$, Fig.~\ref{fig:fig2}~(a)-(b).
Thus, basing only on the mean EE, one might erroneously conclude that the system is ergodic for $D>1/2$, even though the wave-function is confined in an exponentially small ratio of the total Hilbert space $\sim 2^{-(1-D)L}
$.

To understand the above phenomenon, we consider the structure of the reduced density matrix $\rho_A$, Eq.~\eqref{eq:rho_A}, determined by scalar products of the vectors $\psi_M= \lp \psi_{M,1},\ldots, \psi_{M,N_B}\rp$.
For $M\neq M'$, these vectors are independent~\footnote{Here for simplicity we neglect the constraint related to the normalization condition which puts the only condition on $N$ random elements.} and the off-diagonal elements of $\rho_A$ are almost negligible for $D<1$ due to the sparsness of $\psi_M$, which has only $N^D/N
$ fraction of non-zero elements. Instead, the diagonal elements of $\rho_A$ are given by the norms of the vectors $\psi_M$ and cannot be neglected~\footnote{More accurate analysis in Appendices~\ref{App_Sec:rho_structure} and~\ref{Sec:W_sparse} shows that $\rho_A$ has only $\sim O(1)$ non-zero off-diagonal elements in each row as soon as $D<1/2$, while for $D>1/2$ the collective effect of extensive number of off-diagonal elements is smaller than the one of the diagonal ones}.

This analysis can be clearly seen in Fig.~\ref{fig:fig1}, which shows $\rho_A$, $N_A=2^{L/2}$, for a given random configuration of the S-RPS. As one can notice $\rho_A$ is always nearly diagonal. Moreover, for $D>1/2$, an extensive number of off-diagonal elements become non-zero and the diagonal ones are homogeneously distributed with amplitude $\rho^A_{M,M} \sim 2^{-L/2}$, Fig.~\ref{fig:fig1}~(a)-(b).
As soon as $D$ is smaller than $1/2$, only few off-diagonal elements of $\rho_A$ are non-zero, while the distribution of the diagonal ones is bimodal with $\sim 2^{DL}
$ non-zero terms, Fig.~\ref{fig:fig1}~(c).

As the first approximation, the scaling of EE can be estimated considering only diagonal elements of $\rho_A$, $\overline{\mathcal{S}_1}(L/2)~\sim~-\sum_i \rho_{M,M}^A \ln{\rho_{M,M}^A} \sim D_{ent} L/2\ln{2}$, thus obtaining $D_{ent}=1$ for $D<1/2$ and $D_{ent}=2 D$ for $D\geq1/2$. 
We further support the validity of this diagonal approximation in Appendix~\ref{sec:Diag}. In Fig.~\ref{fig:fig2}~(b), we show $D_{ent}
$ both from the EE and its diagonal counterpart
and find the perfect match with the above prediction.

The diagonal approximation has been used to describe thermodynamic entropy out-of-equilibrium~\cite{Polkovnikov_diag-entropy-1,Polkovnikov_diag-entropy-2} and it can be analytically verified in terms of leading scaling behavior.
As only few off-diagonal elements of $\rho_A$ are non-zero (say, $\rho^A_{M,M'}$ for the $M$th row) one can estimate the Schmidt eigenvalues
$\lambda_M$ and $\lambda_{M'}$ by diagonalizing the $2\times2$-matrix
$\begin{pmatrix}
\rho^A_{M,M} & \rho^A_{M,M'}\\
\rho^A_{M',M} & \rho^A_{M',M'}\\
\end{pmatrix}$.
Finally, by the Cauchy-Bunyakovski-Schwarz inequality $|\rho^A_{M,M'}|^2\leq \rho^A_{M,M}\rho^A_{M',M'}$, one concludes that the Schmidt eigenvalues $\lambda_M$ and $\lambda_{M'}$ scale with $N$ as the diagonal elements $\rho^A_{M,M}$, $\rho^A_{M',M'}$ (see Appendix~\ref{App_Sec:sparseness_D<1/2}).
Furthermore, in this leading approximation the mean EE is given by
\be
\overline{\mathcal{S}_1}(L/2) \simeq -\sum_M \overline{\rho^A_{M,M}\ln \rho^A_{M,M}} \sim \overline{\ln N_0},
\ee
where $N_0$ is the number of non-zero diagonal elements $\rho^A_{M,M} = \sum_{m=1}^{N_B} |\psi_{M,m}|^2$~\cite{lnN0-footnote}, which have almost all the same value (see Fig.~\ref{fig:fig1}).

The probability distribution $P(N_0)$ of $N_0$ can be computed combinatorically.
Let $g_M$ be the number of non-zero elements giving contributions to $\rho^A_{M,M}$.
By construction of the S-RPS we have $\sum_M g_M = N^{D}$. Now, $P(N_0)$ is proportional to the product of
the number of combinations $\lp N^D-1\atop N_0-1\rp$ to realize $N_0$ non-zero $g_M>0$
and the number of combinations $\lp N_A\atop N_0\rp$ to place them among $N_A$ values of $1\leq M\leq N_A$.
The typical $N_0
$ is given by the position of the maximum of its probability distribution
\be
N_0^{\text{typ}} = \frac{N_A N^D}{N_A + N^D}\simeq N^{\min(p,D)} \ ,
\ee
confirming the numerical result, Fig.~\ref{fig:fig2},
\be\label{eq:S_1_analytic}
\overline{\mathcal{S}_1}(A) \simeq
\left\{
\begin{array}{ll}
D\ln N, & D<p \\
\ln N_A, & D>p
\end{array}
\right. \ .
\ee

Importantly, the S-RPS do not have any intrinsic locality due to randomly-chosen positions of the non-zero elements.
Thus, Eq.~\eqref{eq:S_1_analytic} gives a natural upper bound for the maximal EE for generic many-body/multifractal wave functions with support set $\sim N^{D_1}
$~\cite{Basis-specific-footnote}.

Now, we further numerically test our main result, Eq.~\eqref{eq:S_1_analytic}, by computing $\mathcal{S}_{1}(A)$ for a different subsystem~$A$.
Figure~\ref{fig:fig21}~(a) shows the slope $D_{ent}$ of $\overline{\mathcal{S}_1}(N_A) \sim D_{ent} \ln{N_A}$ for $N_A = 2^{L/3}$ as a function of the fractal dimension $D$. For $D>1/3$, we have $D_{ent}=1$ and EE shows ergodic behavior. For smaller $D$, $D_{ent}$ deviates from the infinite temperature thermal value, $D_{ent} = 3D$, in agreement with Eq.~\eqref{eq:S_1_analytic}. The difference $\mathcal{S}_{\rm Page}(L/3) - \overline{\mathcal{S}}(A)$ is shown in the inset in Fig.~\ref{fig:fig21}~(a) supporting the convergence of EE to the Page value $\mathcal{S}_{\rm Page}(L/3)$ up to exponentially small corrections in $L$ (as well as $\mathcal{S}_{\rm Page}(L/2)$ in Fig.~\ref{fig:fig2}~(a)).

Furthermore, our results can be generalized also for the Renyi EE, Eq.~\eqref{eq:S_q},
and for genuine multifractal states (see Appendix~\ref{App_Sec:upper_bound}).
Figure~\ref{fig:fig21}~(b) shows $\overline{\mathcal{S}_q}(A) \sim D_{ent}(q) \ln{N_A}$ with $N_A= 2^{L/2}$  for several $q>1$. In agreement with Eq.~\eqref{eq:S_1_analytic}, we obtain $D_{ent}(q) = 1$ for $D>1/2$ and $D_{ent}(q) = 2D$ otherwise.
The $q$-independence of $D_{ent}(q)$ at $q\ge 1$ is an artefact of S-RPS, due to the only fractal dimension $D_q=D$ for $q>0$ characterizing them.

\begin{figure}
\includegraphics[width=1.\columnwidth]{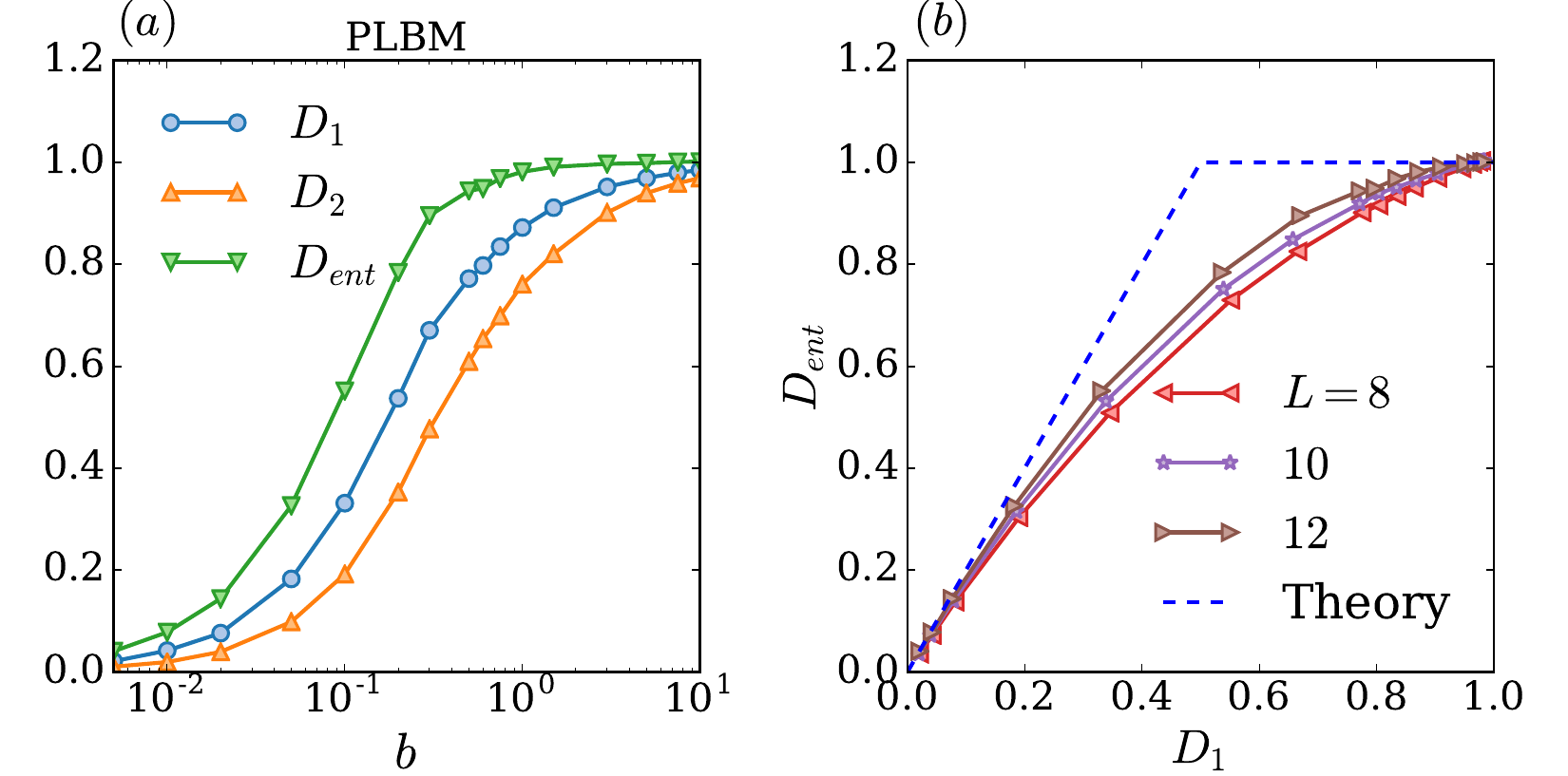}
 \caption{\textbf{Scaling exponents of EE, $D_{ent}$, Shannon entropy, $D_1$, and IPR, $D_2$  for PLRBM.}
 (a)~$D_{ent}$, $D_1$, and $D_2$ extracted by linear extrapolation from
 EE, Shannon entropy $D_1 \ln N$, and the IPR $D_2\ln N$ scalings versus subsystem size $N_A = 2^{L/2}$.
 (b)~The parametric plot of $D_{ent}$ versus $D_1$.
 Different curves correspond to the different points $L-2$, $L$, $L+2$ of an enlarging linear fitting procedure.
 }
 \label{fig:R3}
 \end{figure}
For genuine multifractal states, characterized by non-trivial exponents $D_q$, and for $N_A = N^{p}$, the upper bound, Eq.~\eqref{eq:S_1_analytic}
rewritten as the lower bound for fractal dimensions $\overline{D_{q}}\ln N\geq \mathcal{S}_q(A)$, $q\geq 1$, of a state with fixed Renyi entropies $0\leq \mathcal{S}_q(A)\leq \ln N_A$ can be proved to be strict
and can be saturated by the change of the subsystem bases~\cite{Basis-specific-footnote}.
Indeed, one can show~\cite{MF-footnote} that the minimal value $D_q$
can be achieved if the computational basis is optimized to be
the Schmidt decomposition basis $\left| \psi \right\rangle = \sum_{a} \lambda_{M_a}^{1/2} \left| M_a \right\rangle_A\otimes\left| m_a \right\rangle_B$ as in this case $\psi_{M_a,m_{a'}} = \lambda_{M_a}^{1/2}\delta_{a,a'}$, see Appendix~\ref{App_Sec:upper_bound}.

In order to demonstrate the validity of this general bound in Fig.~\ref{fig:R3}, we show the scaling of the fractal exponents, $D_1$ and $D_2$ extracted from $\text{IPR}_q$, and $D_{ent}$
from EE, for the paradigmatic example of power-law random banded matrices (PLBM)~\cite{Evers2008Review} known to have genuine multifractality of eigestates at the critical point, tuned by the parameter $b$.
Plotting $D_{ent}$ versus $D_1$, we see that the universal bound is satisfied, though not saturated for the inspected system sizes.
We show the results for another exemplary many-body system in Appendix~\ref{App_Sec:num_egs}.
%
%

\textit{Fluctuations}---
Quantum fluctuations represent another important ingredient to understand ergodicity. According to ETH, they can be related to temporal fluctuations around the equilibrium value in a quench protocol.
In ergodic systems the scaling of fluctuations is related to the dimension of the larger subsystem playing the role of a bath~\cite{Majumdar2010RPS,Majumdar2011RPS,Majumdar2015Oxford,Vivo2016RPS}.
EE fluctuations can be quantified by the standard deviation
\be
\overline{\delta \mathcal{S}_1}(A) = ( \overline{\mathcal{S}_1^2}(A) - \overline{\mathcal{S}_1}^2(A))^{1/2} \sim N^{-D_{fluc}/2},
\ee
from the collapse with $L$ of the probability distribution $\mathcal{P}(x)$ of the rescaled variable $x=(\mathcal{S}(A)-\overline{\mathcal{S}}(A))/{\overline{\delta \mathcal{S}}(A)}$ (see Appendix~\ref{App_Sec:P(S)_collapse}).
Importantly, the scaling of fluctuations displays three different regimes for a generic cut $N_A = N^{p}$, $p\leq 1/2$, (see inset in Fig.~\ref{fig:fig1}    for $p=1/2$)
\be\label{eq:S_fluct}
D_{fluc} = \left\{
\begin{array}{ll}
D, & D<p \\
2D-p, & p<D<1-p/2 \\
2(1-p), & D>1-p/2
\end{array}
\right. \ .
\ee
For $D<p$, both mean EE and its fluctuations show the properties of local observables: their scaling is related to the equilibration within the fractal support set $N^D$ and does not depend on the subsystem size.
For $p<D<1-p/2$ the mean EE saturates at the Page value for the considered subsystem size, Eq.~\eqref{eq:S_1_analytic}, and, thus, for such states EE cannot be considered as a local observable. The fluctuations in this case have fingerprints of a non-ergodic behavior, $\overline{\delta \mathcal{S}}(A)\sim N^{-(2D-p)}$. Finally for $1-p/2<D<1$, both mean and its fluctuations are undistinguishable from  ergodic states at infinite temperature~\footnote{Note that unlike the general upper bound for the mean Renyi entropy, the fluctuations are highly system-specific.
For the S-RPS they scale as~\eqref{eq:S_fluct}, see Appendix~\ref{App_Sec:P(S)_collapse}, while in the case of the computational basis being the one of the Schmidt decomposition they are simply zero}.


\textit{Conclusions}---
In order to answer to the main question of the paper~--~to what extend ergodicity properties extracted from entanglement and multifractal probes provide the same description of thermal phases~--~we generalized the result of Ref.~\cite{page1993} on the EE for RPS to the case of NEE states characterized by fractal dimensions $D_q$.
In particular, we presented an upper bound for the entanglement entropies $\mathcal{S}_q$ (both von Neumann and Renyi)
of a generic multifractal state with fractal dimensions $D_q$, see Appendix~\ref{App_Sec:upper_bound}.

This bound leaves the gap for $\mathcal{S}_q(A)$ to be equal to the Page value provided the wave function support set is larger than the subsystem size, $N^{D_q}>N_A$.
An example of the saturation of this bound is shown for a newly introduced class of {\it sparse} random pure states. Our results show that for small fractal dimensions $N^{D_q}<N_A$ EE behaves as a local observable both in terms of the mean value and fluctuations.

Thus, ergodicity viewed as the wave function equipartition in the full Hilbert space is more strict than the one imposed by the value of the EE.

Our results find immediate application in the many-body localization theory where EE is used to probe the transition, or in strongly kinematically constrained models where ergodicity may break down due to Fock/Hilbert space fragmentation.
For instance, in spin models in Refs.~\cite{Sala2019,detomasi2019dynamics}, the eigenstates live on an exponentially small fraction of the full Hilbert, due to dipole conservation~\cite{Sala2019,weak_fragment_footnote} and strong interactions~\cite{detomasi2019dynamics} (Fock-space fragmentation). Nevertheless, the half-chain entanglement entropy equals to the Page value, provided the fractal dimension of the wave function support set is $D>1/2$, see Appendix~\ref{App_Sec:num_egs}.

\begin{acknowledgments}
We thank S. Bera, M. Haque, M. Heyl,  D. Luitz, R. Moessner, F. Pollmann, and S. Warzel for helpful and stimulating discussions.
We are pleased to acknowledge V. E. Kravtsov for motivating us to initiate this project.
We  also  express  our  gratitude  to P. Sala for insightful comments and critical reading of the manuscript.
GDT acknowledges the hospitality of Max Planck Institute for the Physics of Complex Systems Dresden, where part of the work was done.
I.~M.~K. acknowledges the support of German Research Foundation
(DFG) Grant~No.~KH~425/1-1 and the Russian Foundation for Basic Research Grant No. 17-52-12044.
\end{acknowledgments}
\bibliography{Sq_bib}

\begin{thebibliography}{92}
\bibitem [{MF-()}]{MF-footnote}%
  \BibitemOpen
  \href@noop {} {}\bibinfo {note} {The relation between the Renyi entropies
  $S_q$ of genuine multifractal states and their fractal dimensions $D_q$,
  Eq.~\eqref{eq:D_q}, being a non-trivial function of the moment $q$, is
  considered in details by us in~\cite{DeTomasi_MF_Sq} and in Appendix
  G.}\BibitemShut {Stop}%
\bibitem [{lnN()}]{lnN0-footnote}%
  \BibitemOpen
  \href@noop {} {}\bibinfo {note} {As we show in Appendix~E this approximation
  is good for mean EE for any $D$, but fails to capture fluctuations for $D>p$
  due to multifractal distribution of $\rho^A_{M,M}$.}\BibitemShut {Stop}%
\bibitem [{Bas()}]{Basis-specific-footnote}%
  \BibitemOpen
  \href@noop {} {}\bibinfo {note} {Note that both EE and IPR are strongly
  basis-dependent, thus, generic many-body states may present any point under
  the found upper bound by either changing the partition spatial structure
  (changing only $S_q$) or by basis transformations of subsystems (changing
  only $D_q$).}\BibitemShut {Stop}%
\end{thebibliography}

\appendix

%
%

\section{Numerical tests of diagonal approximation}\label{sec:Diag}

\begin{figure*}[t]
 \includegraphics[width=0.9\textwidth]{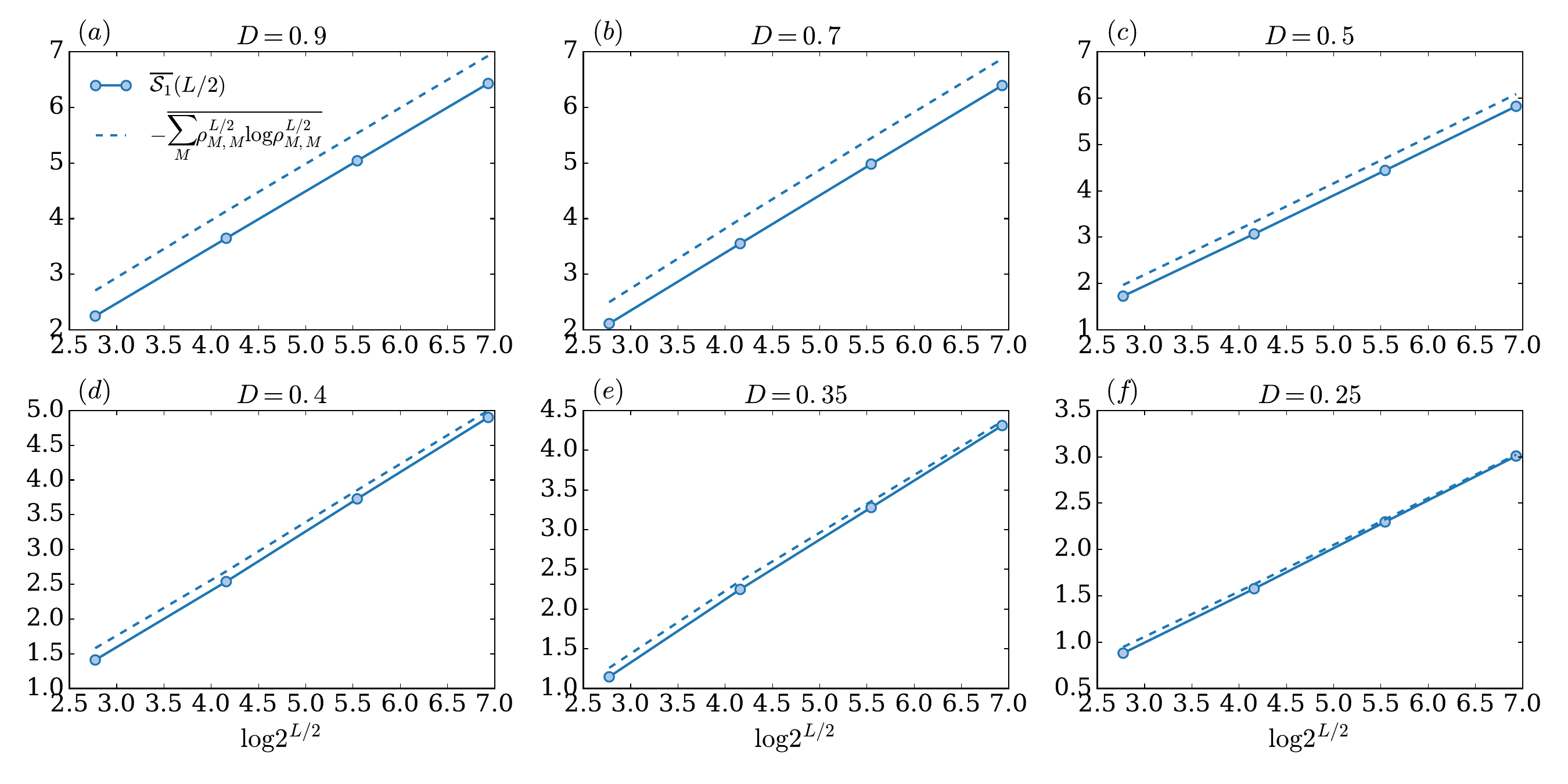}
 \caption{{\bf Diagonal approximation and von Neumann EE} (a)-(f) half-partition averaged EE $\overline{\mathcal{S}_1}(L/2)$ (solid-line) and its diagonal approximation $\overline{\mathcal{S}_1^{\text{diagonal}}}(L/2) = -\overline{\sum_M \rho_{M,M}^{A} \ln{\rho_{M,M}^{A}}}$ (dashed line) as a function of $L$ for several fractal dimension $D$.}
 \label{fig:Sfig1}
\end{figure*}

In this section, we provide further numerical evidence of the validity of the diagonal approximation. In the main text we used the diagonal approximation $\overline{\mathcal{S}_1^{\text{diagonal}}}(A) = -\overline{\sum_M \rho_{M,M}^{A} \ln{\rho_{M,M}^{A}}}$ to estimate the scaling of the von Neumann EE $\mathcal{S}_1(A)$. Figure~\ref{fig:Sfig1} show both $\mathcal{S}_1(A)$ and $\overline{\mathcal{S}_1^{\text{diagonal}}}(A)$ for half-partition $N_A= 2^{L/2}$, as function of $L$ for several fractal dimension $D$, giving indication that  $\overline{\mathcal{S}_1}(A) \simeq \overline{\mathcal{S}_1^{\text{diagonal}}}(A)$.

\begin{figure}[t!]
 \includegraphics[width=0.9\columnwidth]{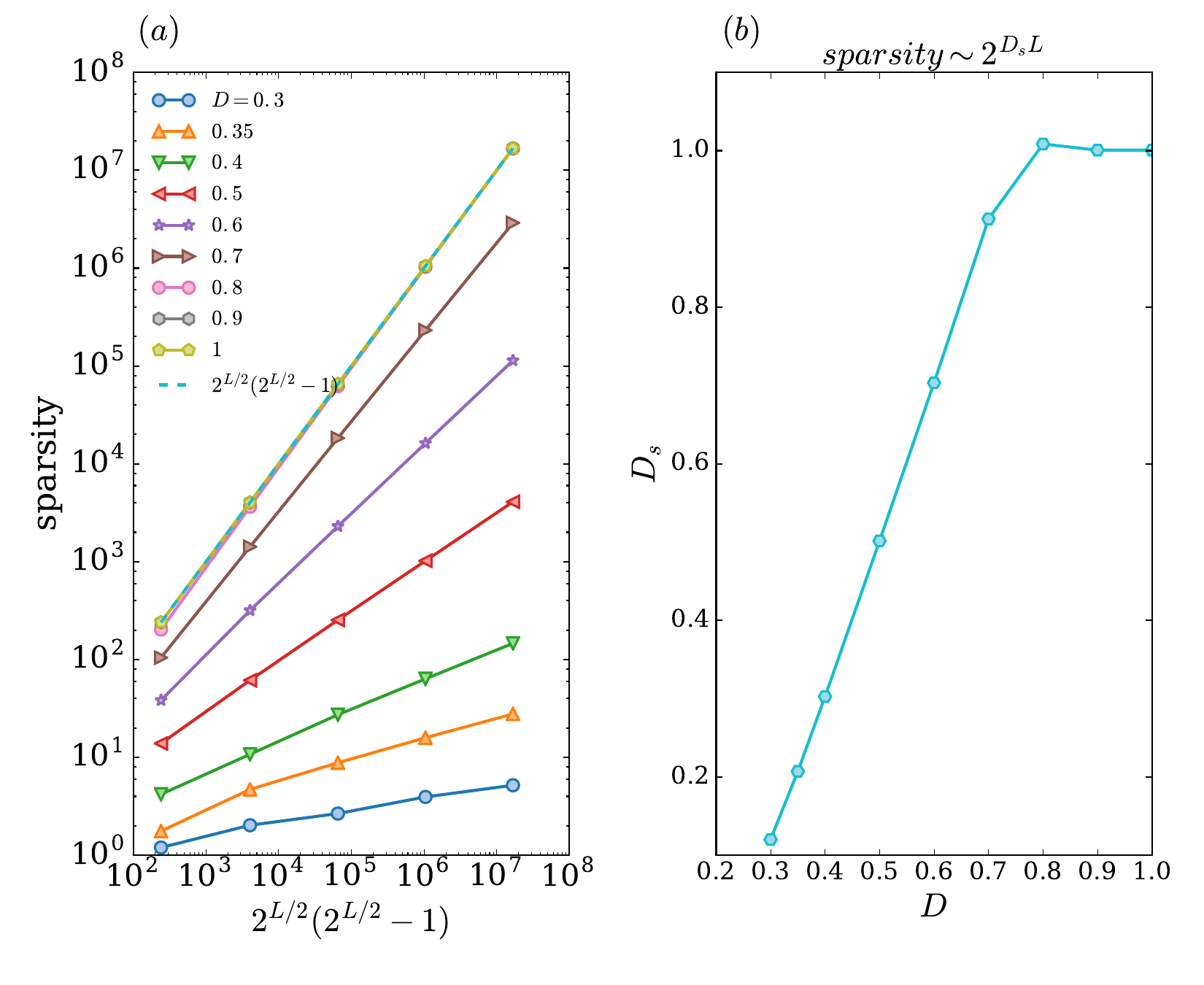}
 \caption{{\bf Sparsity of the off-diagonal elements of $\rho_A$ for $N_A = 2^{L/2}$}.
 (a)~Sparsity $S$ defined as the number of non-zero off-diagonals in the density matrix as a function of the total number $N_A(N_A-1)$ of off-diagonals of $\rho_A$ at half-partition ($N_A=2^{L/2}$) for several $D$. (b)~Rate $D_s$ of non-zero off-diagonal elements, $S\sim 2^{2 p D_s L}$ as a function of the fractal dimension $D$.}
 \label{fig:Sfig2}
 \end{figure}

 \begin{figure}[t!]
 \includegraphics[width=0.9\columnwidth]{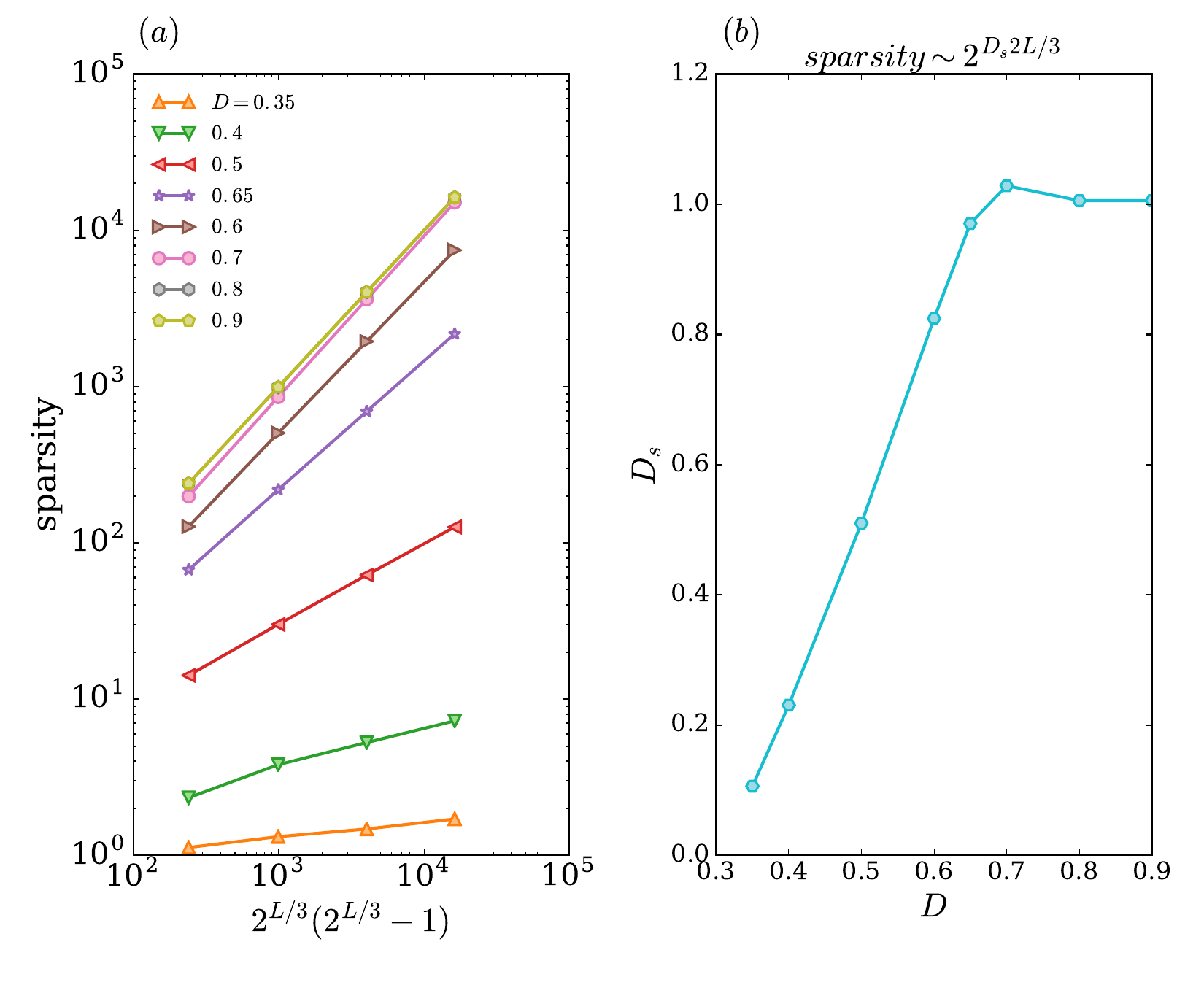}
 \caption{{\bf Sparsity of the off-diagonal  elements of $\rho_A$ for $N_A = 2^{L/3}$}.
 (a)~Sparsity as function of $N_A(N_A-1)$ for several $D$.
 (b)~Rate $D_s$ of non-zero off-diagonal elements, $S\sim 2^{2 p D_s 2 L/3}$ as a function of the fractal dimension $D$.}
 \label{fig:Sfig3}
 \end{figure}

To understand numerically why the diagonal approximation works well, we analyze in more detail the structure of the reduced density matrix $\rho_A$.
First, we start to investigate the sparse property of $\rho_A$. For this purpose, we define the sparsity of $\rho_A$ as the number of its non-zero off-diagonal elements, $S=\# \{ |\rho_A^{M,M'}|\ne0\}$.
Figure~\ref{fig:Sfig2}(a) and Fig.~\ref{fig:Sfig3}(a) show the sparsity of $\rho_A$ for two different partition $N_A=2^{L/2}$ and $N_A=2^{L/3}$, respectively. As one can notice, for large $D$ the number of non-zeros in $\rho_A$ grows as the total size of the reduced density matrix $N_A^2$ meaning that the matrix is not sparse for these values of $D$ . Nevertheless, for small $D$, $\rho_A$ is sparse and the number of non-zeros elements of $\rho_A$ is an exponentially small fraction of the full dimension.

To better quantify the sparsity of $\rho_A$, we define the rate $D_s$ as $S\sim N^{2 D_s p}$. For $D_s=1$ the matrix is not sparse, and $\rho_A$ is diagonal for $D_s=0$. Figure~\ref{fig:Sfig2}(b) and Fig.~\ref{fig:Sfig3}(b) show $D_s$ for two different partitions $N_A=2^{L/2}$ and $N_A=2^{L/3}$, respectively. As expected, for large $D$ we have $D_s=1$, while $D_s$ is proportional to $D$ for smaller $D$.
In the next section, we will give an analytical argument showing
\be\label{eq:Ds_analytic}
D_s \simeq
\left\{
\begin{array}{ll}
\frac{2D-1+p}{2p}, & D<\frac{1+p}{2} \\
1, & D>\frac{1+p}{2}
\end{array}
\right. \ .
\ee
and demonstrate that sparsity plays a major role for the validity of the diagonal approximation.

\begin{figure}[h!]
 \includegraphics[width=0.9\columnwidth]{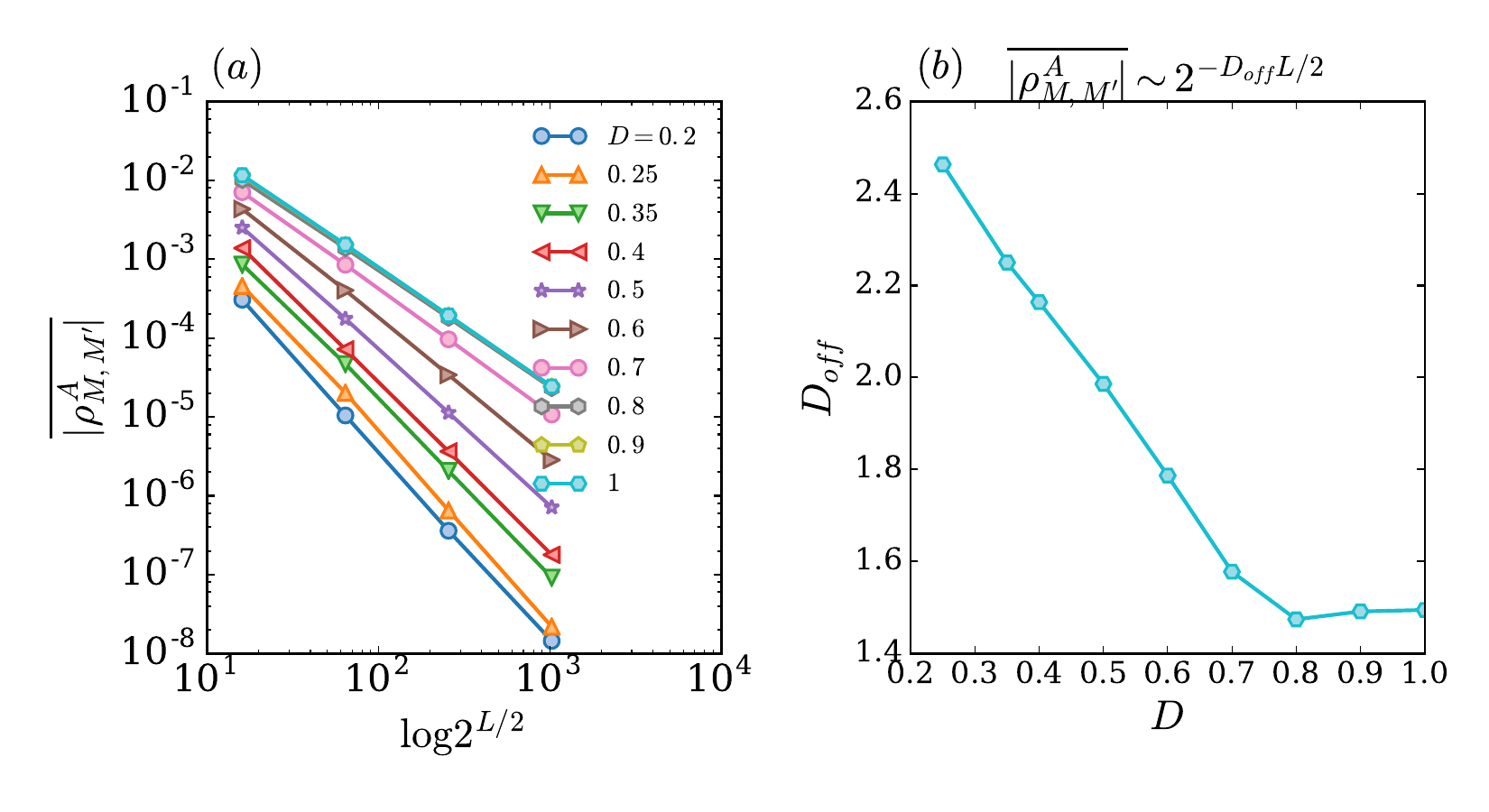}
 \caption{{\bf Mean off-diagonal element of $\rho_A$ for $N_A = 2^{L/2}$}. (a)~ $\overline{|\rho_{M,M'}^A|}$ as a function of $L$ for several $D$. (b)~$D_{off}$ exponent extracted from $\overline{|\rho_{M,M'}^A|}\sim 2^{-p D_{off} L}$, $p=1/2$ as a function of $D$.}
 \label{fig:Sfig4}
 \end{figure}
 \begin{figure}[h!]
 \includegraphics[width=0.9\columnwidth]{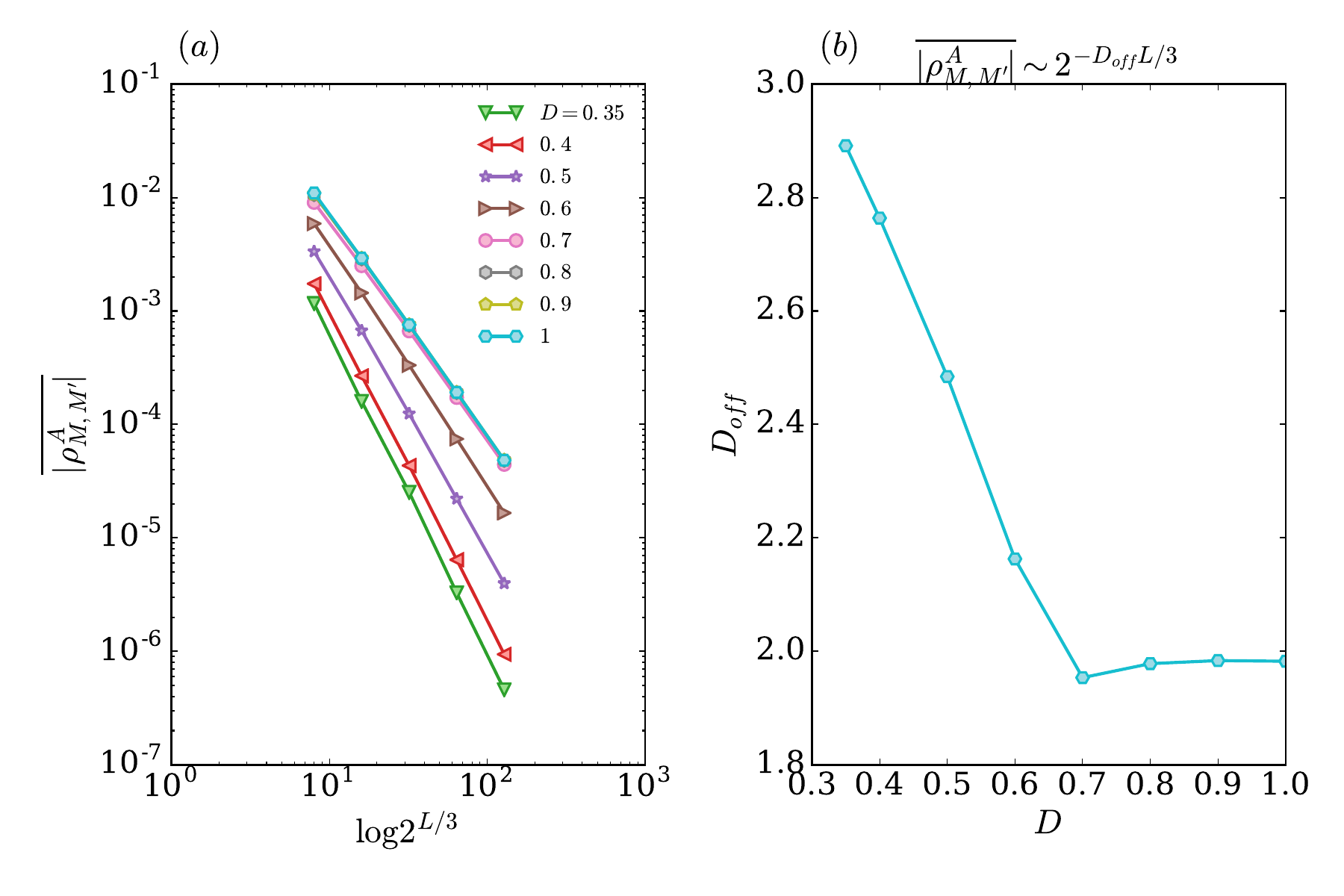}
 \caption{{\bf Mean off-diagonal element of $\rho_A$ for $N_A = 2^{L/3}$}. (a)~ $\overline{|\rho_{M,M'}^A|}$ as a function of $L$ for several $D$. (b)~$D_{off}$ exponent extracted from $\overline{|\rho_{M,M'}^A|}\sim 2^{-p D_{off} L}$, $p=1/3$ as a function of $D$.}
 \label{fig:Sfig5}
 \end{figure}

Now, we calculate the mean off-diagonal elements of $\rho_A$ (not only non-zero ones). Figure~\ref{fig:Sfig4}~(a) and Fig.~\ref{fig:Sfig5}~(a) show $\overline{|\rho_{M,M'}^A|}$ as function of $L$ for several $D$ for two different partitions $N_A=2^{L/2}$ and $N_A=2^{L/3}$, respectively. In general, we have $\overline{|\rho_{M,M'}^A|}\sim N^{-D_{off} p}$.  Figure~\ref{fig:Sfig4}~(b) and Fig.~\ref{fig:Sfig5}~(b) show $D_{off}$ as a function of $D$.
In the next section, we will show that
\be\label{eq:Doff_analytic}
p D_{off} \simeq
\left\{
\begin{array}{ll}
1+p-D, & D<\frac{1+p}{2} \\
\frac{1+p}{2}, & D>\frac{1+p}{2}
\end{array}
\right. \ .
\ee

\section{Structure of reduced density matrix}\label{App_Sec:rho_structure}
In this section, we consider the structure of diagonal
\be\label{eq:rho_diag}
\rho_{M,M}^{A} = \sum_{m=1}^{N_B} |\psi_{M,m}|^2 \ ,
\ee
and off-diagonal
\be\label{eq:rho_offdiag}
\rho_{M,M'}^{A} = \sum_{m=1}^{N_B} \psi_{M,m}\psi^*_{M',m} \ ,
\ee
elements of the reduced density matrix $\rho_A$
assuming the vectors $\psi_M$ and $\psi_{M'}$ to be uncorrelated for $M\ne M'$
with a certain probability distribution of each element
\begin{multline}\label{eq:P(psi)}
P(\psi_{M,m}) = (1-p_0) \delta(\psi_{M,m}) + \\ p_0 P_1(N^{D/2} \psi_{M,m})N^{D/2}  \ .
\end{multline}
Here, $p_0 = N^{D}/N$ is the probability that $\psi_{M,m} \ne 0$. $P_1(y)$ is the probability distribution of non-zero values, which  is symmetric $P_1(-y) = P_1(y)$, has a unit variance
$\int y^2 P_1(y) dy = 1$ and the fourth cumulant $\sigma^2 = \int (y^2-1)^2 P_1(y) dy \sim O(1)$ .
The latter conditions ensure the scaling $|\psi_{M,m}|^2 \sim N^{-D}$ of non-zero elements and the wave function normalization (on average).
In the limit of large $N$, we can further neglect the correlations related to the normalization condition.

Next, within the above assumptions one can find the probability distributions of diagonal, Eq.~\eqref{eq:rho_diag}, and off-diagonal, Eq.~\eqref{eq:rho_offdiag}, elements of the reduced density matrix (similar to~\cite{Behemoths2019}).
For this purpose we rewrite Eq.~\eqref{eq:P(psi)} in a short form for $N^{D/2}\psi_{M,m}=y$
\be\label{eq:P(N^D psi)}
P(y) = (1-p_0) \delta(y) + p_0 P_1(y) \ .
\ee

\subsection{Probability distribution of diagonals $\rho_{M,M}^{A}$}
Here we use the Fourier transform to calculate the $N_B$-fold convolution of the probability distribution $\tilde P_1(t') = \frac{P_1\lp\sqrt{t'}\rp}{\sqrt{t'}}$ of $t' = |\psi_{M,m}|^2$ and obtain
\be\label{eq:P(rho_M)}
P(\rho_{M,M}^{A}) = \sum_{k=0}^{N_B} \lp N_B \atop k\rp (1-p)^{N_B-k} p^k \tilde P_k(\rho_{M,M}^{A} N^D) \ ,
\ee
with
\be
\tilde P_k(t) = \frac{1}{2\pi}\int e^{-i\omega t} \lp \int \frac{P_1\lp\sqrt{t'}\rp}{\sqrt{t'}}e^{i\omega t'}\rp^k d\omega \ .
\ee

The scaling of $p_0 = N^{D-1}$ and $N_B = N^{1-p}$ provide the optimal index
\be
k_* = N_B p_0=N^{D-p}
\ee
giving the main contribution to the sum Eq.~\eqref{eq:P(rho_M)}.

As $k$ is integer, one has to distinguish two cases:
(i)~$D<p$  when $k_*=N_B p_0\ll 1$ and, thus, the probability distribution is nearly bimodal
\begin{multline}\label{eq:P(rho_M)_D<p}
P(\rho_{M,M}^{A}=x)dx \simeq (1-k_*) \delta(x)dx + k_* \tilde P_1(N^{D} x)N^{D} dx \ ,
\end{multline}
and (ii)~$D>p$  when $k_*=N_B p_0\gg 1$ and the central limit theorem (CLT) works giving
\be\label{eq:P(rho_M)_D>p}
P(\rho_{M,M}^{A}) = \frac{e^{-(\rho_{M,M}^{A}-N^{-p})^2/(2\sigma^2 N^{-D-p})}}{\sqrt{2\sigma^2 N^{-D-p}}} \ .
\ee

This analysis shows that for $D>p$ the diagonal $\rho_A$-elements are homogeneously distributed with the mean value $\rho_{M,M}^{A} = 1/N_A$ given by $\text{Tr}[\rho_A] = 1$.

\subsection{Probability distribution of off-diagonals $\rho_{M,M'}^{A}$}
To obtain $P(\rho_{M,M'}^{A})$ one has to calculate, first, from Eq.~\eqref{eq:P(N^D psi)}
\begin{multline}
P(N^D\psi\psi'=z) = \iint P(y)P(y') \delta(z-y y')dy dy' \\ =
(1-p^2) \delta(z) + p_0^2 \bar P(z) \ ,
\end{multline}
with
\be
\bar P_1(z) = \iint P_1(y)P_1(y') \delta(z-y y')dy dy' \ .
\ee
Then, analogously to the previous subsection, one can use the Fourier transform to calculate
\be\label{eq:P(rho_MM')}
P(\rho_{M,M'}^{A}) = \sum_{l=0}^{N_B} \lp N_B \atop l\rp (1-p_0^2)^{N_B-l} p_0^{2l} \bar P_l(\rho_{M,M}^{A} N^D) \ ,
\ee
with
\be
\bar P_l(t) = \frac{1}{2\pi}\int e^{-i\omega t} \lp \int \bar P_1\lp z'\rp e^{i\omega z'}\rp^l d\omega \ .
\ee

The scaling of $p_0 = N^{D-1}$ and $N_B = N^{1-p}$ provide the optimal index
\be
l_* = N_B p_0^2=N^{2D-1-p}
\ee
giving the main contribution to the sum Eq.~\eqref{eq:P(rho_MM')}.

As $l$ is integer, one has to distinguish two cases:
(i)~$D<\frac{1+p}{2}$  when $l_*=N_B p_0^2\ll 1$ and, thus, the probability distribution is nearly bimodal
\begin{multline}\label{eq:P(rho_MM')_D<p}
P(\rho_{M,M'}^{A}=x)dx \simeq (1-l_*) \delta(x)dx + \\ l_* \bar P_1(N^{D} x)N^{D} dx \ ,
\end{multline}
and (ii)~$D>\frac{1+p}{2}$  when $l_*=N_B p_0^2\gg 1$ and CLT works giving
\be\label{eq:P(rho_MM')_D>p}
P(\rho_{M,M'}^{A}) = \frac{e^{-(\rho_{M,M'}^{A})^2/(2\sigma^2 N^{-1-p})}}{\sqrt{2\sigma^2 N^{-1-p}}} \ .
\ee
Here we used the fact that $\bar P_1(z) = \bar P_1(-z)$ is symmetric and thus there is no drift in CLT.

The latter analysis confirms the scaling of the off-diagonal elements, Eq.~\eqref{eq:Doff_analytic}, as well as the number of non-zero off-diagonals, Eq.~\eqref{eq:Ds_analytic}.
Indeed, for $D>\frac{1+p}{2}$ the distribution is smooth with the typical value
\be
\rho_{M,M'}^{A} \sim N^{-\frac{1+p}{2}} \ ,
\ee
thus, the rate of non-zero off-diagonal elements $D_s=1$ and their scaling $\overline{|\rho_{M,M'}^A|}\sim 2^{-p D_{off} L}$ is $D_{off} = \frac{1+p}{2p}$.

In the opposite limit of $D<\frac{1+p}{2}$ the distribution is bimodal giving the number of non-zeros
\be
N^{2 p D_s} = N_A^2 l_* = N^{2D-1+p}
\ee
as well as the mean value
\be
\overline{|\rho_{M,M'}^A|} = N^{-p D_{off}} = l_* N^{-D} = N^{D-1-p} \ .
\ee

\section{Sparseness of the reduced density matrix for non-ergodic states}\label{Sec:W_sparse}

Now, we provide an analytical argument to support the validity of the diagonal approximation in the regime in which $\rho_A$ is sparse.
As we are interested in the scaling of the Schmidt values with $N$ compared to the one of diagonal elements $\rho_{M,M}^{A}$, we have to consider two cases:
(i)~First, when the number of non-zero elements in each row is finite and does not grow with $N$, the off-diagonal elements can be of the same order as the diagonal ones.
(ii)~Second, when there are many non-zero off-diagonals which are much smaller than $\rho_{M,M}^{A}$.

\subsection{Few non-zero off-diagonal elements $\rho_{M,M'}^{A}$, ($D<1/2$)}\label{App_Sec:sparseness_D<1/2}
As follows from Eq.~\eqref{eq:Ds_analytic} there is at most $O(1)$ non-zero off-diagonal elements in each row as soon as $D<1/2$ (the total number of off-diagonals $\sim N_A$).

In this case, we can show that in terms of multifractal scaling with the total Hilbert space dimension $N$ in the above regime
the Schmidt values $\lambda_M$ scale in the same way as the diagonal elements of $\rho_A$ and, thus, EE can be approximated by its diagonal counterpart~\cite{Polkovnikov_diag-entropy-1,Polkovnikov_diag-entropy-2}
\be\label{eq:Sigma_q}
\mc{S}_q(p) = \frac{\ln\Sigma_q}{1-q}, \quad
\Sigma_q=\sum_M \lambda_M^q\simeq\sum_M \lp \rho^A_{M,M}\rp^q \ .
\ee
Indeed, if in each row of $\rho_A$ there are only few significantly non-zero off-diagonal matrix elements (say, for $M$th and $M'$th diagonals), then Schmidt eigenvalues can be approximated by diagonalizing a $2$-by-$2$ matrix
\be
\left(
  \begin{array}{cc}
    \rho^A_{M,M} & \rho^A_{M,M'} \\
    \rho^A_{M',M} & \rho^A_{M',M'} \\
  \end{array}
\right) \ .
\ee
Assuming the following scaling $\rho^A_{M,M} \sim N^{-\alpha_M}$, and $\rho^A_{M,M'} \sim N^{-\beta}$, with $\alpha_M\leq\alpha_{M'}$ without loss of generality, we obtain for the corresponding Schmidt values
\begin{multline}
\lambda_{M/M'} = \frac{N^{-\alpha_M}+N^{-\alpha_{M'}}\pm\sqrt{\lp N^{-\alpha_M}+N^{-\alpha_{M'}}\rp^2+4N^{-2\beta}}}{2} \simeq\\
\left\{
N^{-\alpha_M} + N^{-2\beta+\alpha_M}\simeq N^{-\alpha_M}\atop
N^{-\alpha_{M'}} + N^{-2\beta+\alpha_M}\simeq N^{-\alpha_{M'}}
\right. \ .
\end{multline}
The latter approximation is based on the inequality $\beta\geq(\alpha_M+\alpha_{M'})/2$ leading from
the Cauchy-Bunyakovski-Schwarz inequality for the off-diagonal element by the geometric mean of diagonals
\begin{multline}\label{eq:W_MM'_ineq}
\lv \rho^A_{M,M'}\rv = \lv \sum_{m=1}^{N_B} \psi_{M,m}\psi_{M',m}^*\rv\leq \\
\sqrt{\sum_{m=1}^{N_B} \lv\psi_{M,m}\rv^2 \sum_{m'=0}^{N_B} \lv\psi_{M',m'}\rv^2} = \sqrt{\rho^A_{M,M} \rho^A_{M',M'}} \ .
\end{multline}
As a result, the scaling of Schmidt values $\lambda_M$ with $N$ is shown to be the same as for the diagonal elements $\rho^A_{M,M}$
in the nearly diagonal sparse regime of $\rho_A$ ($D<1/2$).
In next sections we will use this fact to calculate the Renyi and entanglement entropies.

\subsection{Many non-zero off-diagonal elements $\rho_{M,M'}^{A}$, ($D>1/2$)}
In the case of $D>1/2$ there is an extensive number of non-zero off-diagonal elements of the reduced density matrix.
In order to estimate them we assume their statistical independence from each other and from the diagonal elements following the case $D=1$ considered in~\cite{Vivo2016RPS}.

In the case of $D>\frac{1+p}2>p$ both diagonal and off-diagonal elements of
the reduced density matrix are homogeneously distributed and the latter has the form similar to the Rosenzweig-Porter random matrix ensemble~\cite{RP}.
Then the Schmidt eigenspectrum is not affected by the off-diagonal elements when~\cite{ShapiroKunz}
\be
\frac{|\rho_{M,M'}^{A}|}{\rho_{M,M}^{A}}\sim N^{-\frac{1-p}{2}}\ll N^{-p/2} \ ,
\ee
which is the case as $p\leq 1/2$.

In general for $D>1/2$ one can apply the Mott's principle of delocalization~\cite{Mott1966} recently generalized in~\cite{Nosov2019correlation} which reads for $N_A\times N_A$ matrix, $N_A = N^p$, as follows: the spectrum is not affected by the off-diagonal elements as soon as
\be
N^p\frac{\overline{|\rho_{M,M'}^{A}|^2}}{\overline{|\rho_{M,M}^{A}|^2}}\ll 1 \ .
\ee
In our case it leads to
\be
N^p\frac{\overline{|\rho_{M,M'}^{A}|^2}}{\overline{|\rho_{M,M}^{A}|^2}}\sim 
\left\{
\begin{array}{ll}
N^{2D-2+p}, & D<\frac{1+p}{2} \\
N^{2p-1}, & D>\frac{1+p}{2}\end{array}
\right.
 \ ,
\ee
and works for any $0\leq D\leq 1$.

\section{Entanglement entropy for fractal states}
In this section, we consider mean and fluctuations of Renyi and von Neumann EE within the approximations of two previous sections.

The simplest way to calculate the Renyi entropy, Eq.~\eqref{eq:Sigma_q},
in the diagonal approximation
\be
\lambda_M\simeq \rho^A_{M,M} = \sum_{m=1}^{N_B} \lv\psi_{M,m}\rv^2
\ee
is to use the probability distributions, Eq.~\eqref{eq:P(rho_M)_D<p}, and, Eq.~\eqref{eq:P(rho_M)_D>p}.
Indeed,
\be
\overline{\Sigma_q} = N_A \overline{\lp\rho_{M,M}^A\rp^q} =
\left\{
\begin{array}{ll}
N^{D(1-q)}, & D<p \\
N^{p(1-q)}, & D>p
\end{array}
\right. \ ,
\ee
leading straightforwardly to Eq.~(8) of the main text.

The fluctuations can be also estimates from the moments as soon as the variance
\be
\sigma_\Sigma^2=\overline{\Sigma_q^2}-\overline{\Sigma_q}^2 = N_A \lb\overline{\lp\rho_{M,M}^A\rp^{2q}} - \overline{\lp\rho_{M,M}^A\rp^{q}}^2\rb
\ee
is small compared to $\overline{\Sigma_q}^2$.
Indeed, as
\begin{multline}
(1-q) \mc{S}_q(p) = \ln\Sigma_q = \ln\overline{\Sigma_q} + \ln\lb 1 + \frac{\sigma_\Sigma}{\overline{\Sigma_q}}g_q\rb \simeq \\
\ln\overline{\Sigma_q} + \frac{\sigma_\Sigma}{\overline{\Sigma_q}}g_q - \frac{\sigma_\Sigma^2}{2\overline{\Sigma_q}^2}g_q^2 \
\end{multline}
it gives
\be
(1-q) \overline{\mc{S}_q(p)} = \ln\overline{\Sigma_q}- \frac{\sigma_\Sigma^2}{2\overline{\Sigma_q}^2}\simeq \ln\overline{\Sigma_q}
\ee
within the leading approximation, and
\be
(1-q)^2 \lb\overline{\mc{S}_q^2(p)}-\overline{\mc{S}_q(p)}^2\rb = \frac{\sigma_\Sigma^2}{\overline{\Sigma_q}^2} \ .
\ee
Here we introduced dimensionless variable $g_q = \frac{\Sigma_q-\overline{\Sigma_q}}{\sigma_\Sigma}$ with zero mean and unit variance
\be
\overline{g_q} = 0 \ , \quad \overline{g_q^2} = 1 \ .
\ee

In our case one obtains
\be
\frac{\sigma_\Sigma^2}{\overline{\Sigma_q}^2}=
N^{-D} \ ,
\ee
giving the correct approximation for $D<p$.

\subsection{Alternative way to calculate entanglement entropies}
Alternatively in the main text we parameterize Schmidt values as follows
\be
\lambda_M\simeq \rho^A_{M,M} = \sum_{m=1}^{N_B} \lv\psi_{M,m}\rv^2 = g_M/N^D \ ,
\ee
where $0\leq g_M\leq N^D$ are integer values summed to the support set $N^D$:
\be\label{eq:g_M_norm}
\sum_{M=1}^{N_A} g_M = N^D \ .
\ee

The entanglement entropy in this case can be estimated as
the logarithm of the number $N_0$ of non-zero $g_M$
\be\label{eq:S_lnN0}
{\mc S}_1 \sim \ln N_0 \ .
\ee
As we show below, this approximation is good for mean EE for any $D$, but fails to capture fluctuations for $D>p$.

The probability distribution $P_{N_0}$ of $N_0$ can be calculated combinatorically in the assumption of homogeneous distribution of $g_M$'s.
Indeed, the total number of combinations of $N_A$ values of $g_M$, $1\leq M\leq N_A$, taken with repetitions ($g_M$ can be larger than $1$) and with the normalization Eq.~\eqref{eq:g_M_norm} is given by
\be
{\mc M} = \lp N_A+N^D-1\atop N^D\rp \ .
\ee
At the same time the combinations with $N_0$ non-zero $g_M$ can be counted as the number of
combination to realize $N_0$ non-zeros
\be
{\mc M}_{\bar0} = \lp N^D-1\atop N_0-1\rp
\ee
times the number of combinations to place them among $N_A$, which is
\be
{\mc M}_{N_0} = \lp N_A\atop N_0\rp \ .
\ee
As a result
\be\label{eq:P_N0}
P_{N_0} = \frac{{\mc M}_{\bar0} {\mc M}_{N_0}}{{\mc M}}\simeq A(N) e^{N_A f(\rho)} \ ,
\ee
where
\be\label{eq:f(rho)}
f(\rho,\alpha) = -2\rho\ln(\rho)-(\alpha-\rho)\ln(\alpha-\rho)-(1-\rho)\ln(1-\rho) \ ,
\ee
$N^D = \alpha N_A$, and $0\leq \rho = N_0/N_A \leq 1,\alpha$ and we neglected $-1$ comparing both to $N^D$ and $N_0$.
The expression for $f(\rho)$ is calculated in the large-$N$ limit with help of Stirling's approximation.

The maximum of $f(\rho)$ is achieved at the typical $N_0^* = N_A \rho^*$ with
\be~
\rho^* = \frac{\alpha}{1+\alpha}<1,\alpha \text{ leading to } N_0^* = \frac{N_A N^D}{N_A+N^D} \
\ee
from the main text.

The relative fluctuations $\delta N_0/{N_0^*}=\delta \rho/{\rho^*}$ can be written in the following form
\be
\frac{\delta N_0}{N_0^*} = \frac{\delta \rho_0}{\rho_0^*} = \frac{1}{\sqrt{-N_A f''(\rho^*)}\rho_0^*} = \lp N_A+N^D\rp^{-1/2} \
\ee
in the Gaussian approximation
\be
P_{N_0} = \frac{e^{-(N_A+N^D)(\rho-\rho^*)^2/2}}{\sqrt{2\pi/(N_A+N^D)}} \ ,
\ee
derived from Eq.~\eqref{eq:P_N0} and Eq.~\eqref{eq:f(rho)} provided $\rho^*\gg (N_A+N^D)^{-1/2}$.

In the same approximation
\bes
\begin{align}\label{eq:ln_rho}
\overline{\ln N_0} = \ln N_0^*-\frac{1}{2(N_A+N^D)}, \\
\label{eq:ln2_rho}
\overline{\ln^2N_0} = \ln^2N_0^*+\frac{1-\ln N_0^*}{(N_A+N^D)} \ .
\end{align}
\ees

According to Eq.~\eqref{eq:S_lnN0} and Eq.~\eqref{eq:ln_rho} mean EE is given by 
\begin{multline}\label{eq:S_1_RP_res}
\overline{\mc S_1} \simeq
\overline{\ln N_0}=
\ln N_0^* - \frac{1}{2(N^D+N_A)} \\
\sim \left\{
\ln N^D = D \ln N, \text{ for } D_1<p\atop
\ln N_A = p \ln N, \text{ for } D_1>p
\right. \ .
\end{multline}
In the latter equality we neglected subleading terms.

At the same time according to Eq.~\eqref{eq:ln_rho} and Eq.~\eqref{eq:ln2_rho} EE fluctuations are given mostly by the relative fluctuations of $N_0$
\be
\overline{\mathcal{S}_1^2} - \overline{\mathcal{S}_1}^2 \simeq
\overline{\ln^2(N_0)}-\overline{\ln(N_0)}^2 \simeq \frac{1}{N^D+N_A} \ .
\ee

As mentioned above the approximation Eq.~\eqref{eq:S_lnN0} works both for the mean and fluctuations provided $D<p$.
This is the case as well for all Renyi entropies.
It is caused by the fact that $N_0\simeq N^D$ and, thus, all $g_M\sim O(1)$ leading to
\be
\overline{\Sigma_q} = \sum_M \overline{\lambda_M^q} = \sum_M \overline{\lp\frac{g_M}{N^D}\rp^q} \sim N_0 N^{-Dq} = N^{D(1-q)}
\ee
and
\be
\overline{\Sigma_q^2}-\overline{\Sigma_q}^2 = \sum_M \overline{\lp\frac{g_M}{N^D}\rp^{2q}}-\overline{\lp\frac{g_M}{N^D}\rp^{q}}^2
\sim N^{D(1-2q)} \ .
\ee

However, in the opposite case $D>p$ when $N_0\simeq N_A\ll N^D$ there is a non-trivial distribution of $g_M$ with $\overline{g_M} = N^D/N_A\gg 1$ and
\be
\ln N_0 \neq \frac{\ln\lb\sum_M \lp\frac{g_M}{N^D}\rp^q\rb}{1-q} \ .
\ee
Nevertheless, as we have shown in Sec.~\ref{sec:Diag}, on average both sides of the latter equation give the same Page value as Eq. (8) in the main text.

\section{Collapse of the EE probability distribution for S-RPS}\label{App_Sec:P(S)_collapse}
\begin{figure*}
 \includegraphics[width=1.\textwidth]{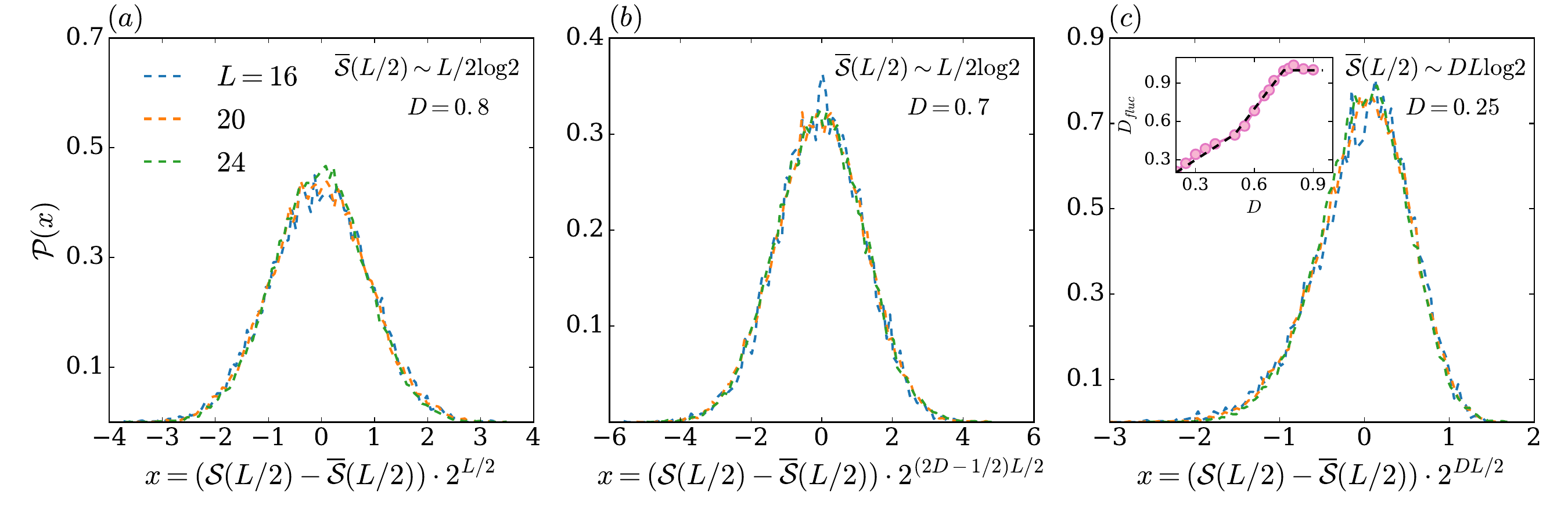}
 \caption{{\bf Collapse of the probability distribution $\mathcal{P}(x)$ of the half-system EE at finite sizes $L$}, $x=(\mathcal{S}-\overline{\mathcal{S}})/{\overline{\delta \mathcal{S}}}$, for different $D$:
 (a)~$D=0.8$, (b)~$D=0.7$, (c)~$D=0.25$.
 Scaling of the mean EE, $\overline{\mathcal{S}_1}(A)$, is shown in the legend, while the distribution width scaling is specified in the axis label.
 (inset)~Slope $D_{fluc}$ of the standard deviation $\overline{\delta \mathcal{S}}(L/2) \sim D_{fluc} \ln{2^{L/2}}$ versus $D$.
 Black dashed line shows analytical prediction, Eq.~\eqref{eq:S_fluct}, for $p=1/2$.
%
}
 \label{fig:fig3}
 \end{figure*}
Here, we numerically characterize EE fluctuations for the S-RPS.
EE fluctuations can be quantified by the standard deviation
\be
\overline{\delta \mathcal{S}_1}(A) = ( \overline{\mathcal{S}_1^2}(A) - \overline{\mathcal{S}_1}^2(A))^{1/2} \sim N^{-D_{fluc}/2},
\ee
from the collapse with $L$ of the probability distribution $\mathcal{P}(x)$ of the rescaled variable $x=(\mathcal{S}(A)-\overline{\mathcal{S}}(A))/{\overline{\delta \mathcal{S}}(A)}$.
Figure~\ref{fig:fig3} shows the collapse of $\mathcal{P}(x)$ with $L$ for S-RPS for several $D$ and $N_A=2^{L/2}$.
Scaling of fluctuations extracted from the above collapse
displays three different kinds of behavior for a generic cut $N_A = N^{p}$, $p\leq 1/2$, (see inset in Fig.~\ref{fig:fig3} for $p=1/2$)
\be\label{eq:S_fluct}
D_{fluc} = \left\{
\begin{array}{ll}
D, & D<p \\
2D-p, & p<D<1-p/2 \\
2(1-p), & D>1-p/2
\end{array}
\right. \ .
\ee
For $D<p$, both mean EE and its fluctuations show the properties of local observables: their scaling is related to the equilibration within the fractal support set $N^D$ and does not depend on the subsystem size~--~and correspond to the analytical results of the previous section.
For $p<D<1-p/2$ the mean EE saturates at the Page value for the considered subsystem size, Eq.~\eqref{eq:S_1_RP_res}, and, thus, for such states EE cannot be considered as a local observable. The fluctuations in this case have fingerprints of a non-ergodic behavior, $\overline{\delta \mathcal{S}}(A)\sim N^{-(2D-p)}$. Finally for $1-p/2<D<1$, both mean and its fluctuations are undistinguishable from  ergodic states at infinite temperature.

\section{Upper bound for the mean entanglement entropy for multifractal states}\label{App_Sec:upper_bound}
For genuine multifractal states the above mentioned derivations are not applicable as its probability distribution is not bimodal like in Eq.~\eqref{eq:P(psi)} or Eq.~\eqref{eq:P(N^D psi)}.
Thus, in order to proceed we use another approach and show that there is a generic upper bound for the mean Renyi entanglement entropy $\overline{\mathcal{S}_q}(A)$ for a state with certain fractal dimensions $0\leq D_q \leq 1$
\be\label{eq:S_q_analytic}
\overline{\mathcal{S}_q}(A) \leq
\left\{
\begin{array}{ll}
D_q\ln N, & D_q<p \\
\ln N_A, & D_q>p
\end{array}
\right. \
\ee
or vice-a-versa there is a generic lower bound for the fractal dimensions $\overline{D_{q}}$ for a state with certain fixed Renyi entropies $0\leq \mathcal{S}_q(A)\leq \ln N_A$
\be\label{eq:D_q_lower_bound}
\overline{D_q}\geq \frac{\mathcal{S}_q(A)}{\ln N} \ .
\ee
Here and further, we restrict our consideration to the physically relevant moments $q\geq 1$.

One should note that both EE and IPR are strongly basis-dependent, thus, generic many-body states may present any point under (above) the found upper (lower) bound.
Indeed, one can see it simply by either changing the partition spatial structure (which changes only $\mathcal{S}_q(A)$, but not $D_q$) or by basis transformations of subsystems (which changes only $D_q$, but not $\mathcal{S}_q(A)$).

As both Eq.~\eqref{eq:S_q_analytic} and \eqref{eq:D_q_lower_bound} are equivalent, let's prove the second one.
For this we consider the state
\be
\lv \psi \ra = \sum_{a} \lambda_{\alpha}^{1/2} \lv E_\alpha \ra_A\otimes\lv \ep_\alpha \ra_B
\ee
written in the Schmidt decomposition basis, where $\lambda_{\alpha}$ are (non-zero) Schmidt eigenvalues and $\lv E_\alpha \ra_A$ and $\lv \ep_\alpha \ra_B$ are the corresponding eigenvectors in subsystems A and B.
From Eq.~\eqref{eq:Sigma_q} the $q$th Renyi entropy is given by
\be
\mc{S}_q(A) = \frac{\ln\lb \sum_M \lambda_M^q\rb}{1-q} \
\ee
and does not depend on the eigenvectors $\lv E_\alpha \ra_A$ and $\lv \ep_\alpha \ra_B$.

On the other hand, the coefficients $\psi_{M,m}$ of $\lv \psi \ra$ in the computational basis $\lv M \ra_A\otimes\lv m \ra_B$ are highly sensitive to these vectors
\be
\psi_{M,m} = \sum_{a} \lambda_{\alpha}^{1/2} \la M| E_\alpha \ra_A \la m| \ep_\alpha \ra_B \ ,
\ee
so are the fractal dimensions $D_q$, Eq.~\eqref{eq:D_q}
\be
{\rm IPR}_q\equiv N^{-D_q (q-1)} = \sum_{M,m}|\psi_{M,m}|^{2q}  \ .
\ee

The minimization of $D_q$ is equivalent to the maximization of the corresponding IPR.
Due to $q>1$ the latter always increases if one replaces
several squared coefficients $|\psi_{M,m}|^{2}$ by their squared sum
due to the generalized Cauchy-Bunyakovski-Schwarz inequality
\be
a^q +b^q < (a+b)^q \ , \quad a,b>0 \ .
\ee
This process is equivalent to the change of the computational basis with respect to the Schmidt basis and
is limited by the fixed Schmidt eigenvalues.

The maximal value of the IPR$_q$ corresponds to the choice of the computational basis to be the same as the Schmidt eigenbasis
\be
\lv M \ra = \lv E_M \ra, \quad \lv m \ra = \lv \ep_m \ra \ .
\ee
and leads to the diagonal reduced density matrix.

The latter gives immediately
\be
\psi_{M,m} = \lambda_{M}^{1/2}\delta_{M,m}
\ee
and
\be
N^{-D_q (q-1)} = \sum_{M,m}|\psi_{M,m}|^{2q} = \sum_M \lambda_M^q = e^{-\mc{S}_q(q-1)}  \ .
\ee
This concludes the proof of the inequality~\eqref{eq:D_q_lower_bound} as in all steps we have just increased the IPR, i.e., decreased $D_q$.

Thus, in a general case of a genuine many-body multifractal state, the inequalities~\eqref{eq:S_q_analytic} and~\eqref{eq:D_q_lower_bound} are valid, however their saturation to equalities is an open and difficult question needed further investigations.

In the main text we have provided an example of S-RPS saturating this bound.
Here as a counterexample let's consider fractal NEE states with the structure of the wave function forming a so-called ergodic bubble~\cite{DeRoeck_PRB_2017,DeRoeck_PRL_2017,DeRoeck_arx_2017,DeRoeck_PRL_2018,Goremykina_PRL_2019,Goremykina_PRB_2019}.
Let's consider the states with only $N^D$ non-zero elements, which are Gaussian distributed with the width controlled by the normalization, but
unlike S-RPS the distribution of these coefficients will be different.
We associate each configuration among $N = 2^L$ ones with the state of $L$ spins-$1/2$ and assume $D L$ to be an integer number.
Then by the ergodic-bubble structure we would mean that
for the randomly chosen $D L$ spins from $L$
the wave-function coefficients are non-zero for all the $2^{D L}$ configurations of the selected spins, while the rest $(1-D) L$ spins are frozen in a fixed state.

It is easy to show that as soon as $D L$ spins are chosen randomly among the subsystems A (with $p L$ spins) and B (with $(1-p) L$ spins)
the fraction of the chosen spins in the subsystem A is the same as in the total system and equal to $D$.
As a result, the reduced density matrix in the subsystem A is represented by the product state of $\sim (1-D) p L$ spins and an ergodic bubble of the size $\sim D p L$.
The resulting entanglement entropy is equal to
\be
\mc{S}_q^{\rm bubble}(A) = D \ln N_A \ ,
\ee
which saturates the bound~\eqref{eq:S_q_analytic} only in the trivial cases $D = 0$ and $D = 1$.


\section{Further numerical examples}\label{App_Sec:num_egs}
In this section we provide two additional examples of the random matrix and many-body models confirming the general results for the upper bound of EE, Eq. (8) of the main text, and the possibility of EE to be saturated at the Page value for strictly non-ergodic wave functions with $D_1<1$.

\subsection{Power-law random banded matrix }
\begin{figure*}[t]
\includegraphics[width=0.9\textwidth]{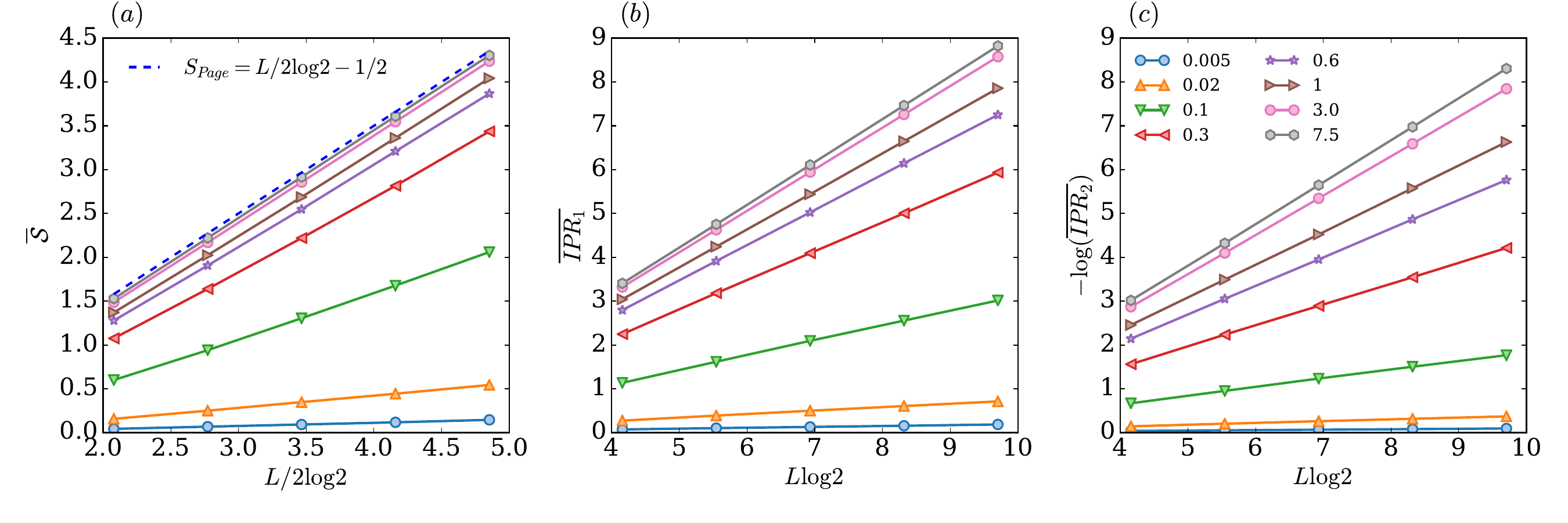}
 \caption{\textbf{Scaling of EE, Shannon entropy $D_1 \ln N$, and the logarithm of IPR $D_2\ln N$ as a function of subsystem size $N_A = 2^{L/2}$ for PLRBM at the critical point, $\alpha=1$ for different multifractal parameter $b$.}
 (a)~Scaling of the mean von Neumann entanglement entropy $\overline{\mathcal{S}_{1}}(L/2)$ of half-system, $N_A=N^{1/2} = 2^{L/2}$, as a function of $L$ . Dashed line stands for the Page value scaling;
 (b)~The mean Shannon entropy $\overline{D_1} \ln N$ versus $L$;
 (c)~The mean IPR fractal dimension $\overline{D_2} \ln N = \ln IPR_2$ versus $L$.
 Values of the multifractal parameter $b$ for all three panels are shown in the legend of panel~(c).
 }
 \label{fig:R2}
 \end{figure*}

\begin{figure}
\includegraphics[width=1.\columnwidth]{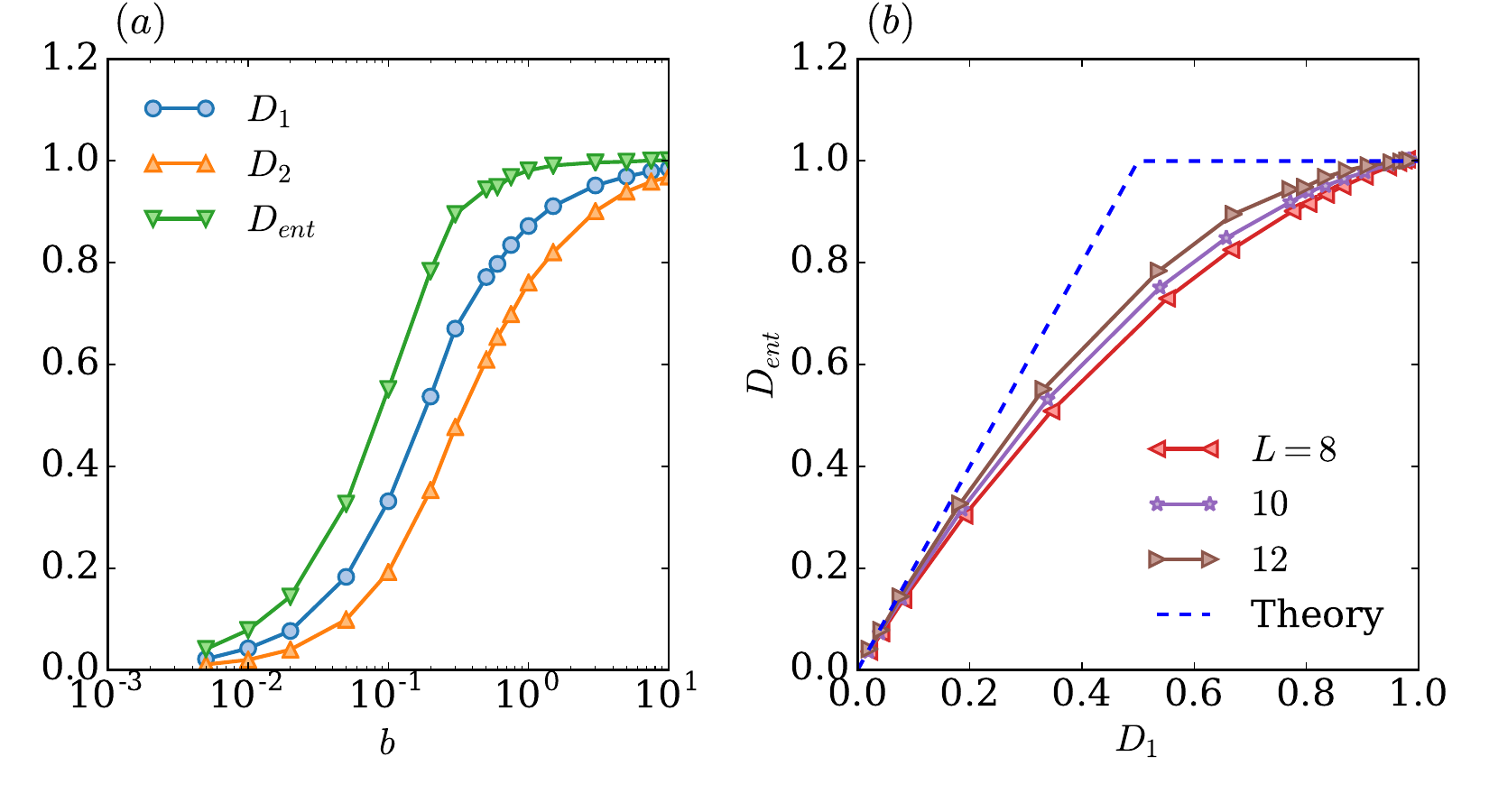}
 \caption{\textbf{Scaling exponents of EE, $D_{ent}$, Shannon entropy, $D_1$, and IPR, $D_2$.}
 (a)~$D_{ent}$, $D_1$, and $D_2$ extracted from Fig.~\ref{fig:R2} by linear extrapolation.
 (b)~The parametric plot of $D_{ent}$ versus $D_1$.
 Different curves correspond to the different points $L-2$, $L$, $L+2$ of an enlarging linear fitting procedure.
 }
 \label{fig:R3}
 \end{figure}
In this section, we focus on entanglement properties of the power-law random banded matrix (PLRBM) ensemble~\cite{Evers2008Review} and their relations with the fractal exponents.
The PLRBM are defined by
\begin{equation}
 \hat H = \sum_{x,y}^{2^L} h_{x,y} |x\rangle \langle y|,
\end{equation}
where $h_{x,y} = \mu_{x,y}/(1 + (|x-y|/b)^{2\alpha})^{1/2}$ and $\mu_{x,y} = \mu_{y,x}$ are independent random variables uniformly distributed in the interval $[-1,1]$.
We consider the case in which the eigenstates of $\hat H$ are multifractal, therefore we fix the parameter $\alpha$ to its critical value at the Anderson localization transition (ALT) $\alpha = 1$~\footnote{Note that recently it has been shown that the delocalization transition in such models crucially depends on the correlations in $\mu_{x,y}$~\cite{Nosov2019correlation} and eigenstates may even be localized beyond the locator expansion convergence~\cite{Deng2018duality,Nosov2019correlation,Kutlin2020_PLE-RG,deng2020}}.
The parameter $b$ tunes the multifractal properties of the PLRBM. For $b \gg 1$ the system is weakly multifractal, meaning that $1 - D_1 \propto b^{-1}$ and for $b\ll 1$ it is in a regime of strong multifractality, $D_1 \propto b$~\cite{Evers2008Review,Mirlin96, Mirlin00,Mirlin93}. $\hat H$ is defined in a Hilbert space of the dimension $2^L$, thus the states $|x\rangle$ could be represented in terms of spin-$1/2$ configurations if one writes them in a binary representation.

Figure~\ref{fig:R2}~(a)  shows the bipartite entanglement entropy $\mathcal{S}$ with the subsystem Hilbert space dimension $N_A = 2^{L/2}$ as a function of $L$ for several $b$. We averaged $\mathcal{S}$ over random configurations, few eigenstates in the middle of the spectrum of $\hat H$ and random partitions.
As expected for large $b$, we have $\mathcal{S} \simeq \mathcal{S}_{Page} = L/2 \log{2}-1/2$. As the parameter $b$ decreases, $\mathcal{S}$ scales still with a volume law, but for $b\lesssim 0.6$ it scales slower than $S_{Page}$, meaning $\overline{\mathcal{S}} \sim D_{ent} L/2 \log{2}$ with $D_{ent}<1$. Figures~\ref{fig:R2}~(b)-(c) show the fractal dimensions $\overline{D_1}$ and $\overline{D_2}$ corresponding to the Shannon entropy and the logarithm of the IPR versus $L$ for several $b$. Usually, the computation of fractal exponents $D_q$ is a challenging task and in principle they could have severe finite-size effects (see, e.g., hot debates on the existence of the non-ergodic extended phase in the Anderson model on the random regular graph).

We estimate the three exponents $D_{ent}$, $D_1$ and $D_2$ by a linear fit repeating Fig.~4 of the main text as Fig.~\ref{fig:R3}. First, Fig.~\ref{fig:R3}~(a) gives numerical evidence of the existence of a genuine multifractal phase, meaning that $D_q$ is a non-trivial function of $q$ as $D_2<D_1$. Second,  as one can notice, there is a regime in $b$ for which $D_{ent}\approx 1$, even tough $D_1<1$. To understand the finite-size effects for $D_{ent}$ and $D_{1}$, we use an enlarging linear fitting procedure. This means that we compute $D_1$ and $D_{ent}$ fitting three consecutive system sizes $\{L-2, L, L+2\}$. Figure~\ref{fig:R3}~(b) shows the result of such fitting $D_{ent}$ as function of $D_1$ for several $L$. As expected, $D_{ent}$ obeys the upper-bound that we provided in the main text (dashed line in Fig.~\ref{fig:R3}~(b)). Moreover, the finite-size flow direction suggests that the upper-bound might be saturated in the thermodynamic limit ($L\rightarrow \infty$).

\subsection{A quantum many-body system}

In conclusion on the main text, we claimed that the results for $D_{ent}$ find an immediate applications for quantum system where Fock/Hilbert space fragmentation take place. In this section, we will give a numerical example for the later model.

We study the $t{-}V$ disordered chain of spinless fermions with periodic boundary conditions,
\begin{equation}
\label{eq:full_H}
 \hat H = -t \sum_x \hat c_{x+1}^\dagger \hat c_{x} + \textit{h.c.} + W\sum_x \mu_x \hat n_x + V\sum_x \hat n_x \hat n_{x+1},
\end{equation}
where $\hat c^\dagger_x$ ($\hat c_x$) is the fermionic creation (annihilation) operator at site $x$, $\hat n_x = \hat c_x^\dagger \hat c_x$, and $\mu_x$ are independent  random  variables  uniformly  distributed in the interval $[-1,1]$. $t=1/2$, $W§$ and $V$ are  the hopping, disorder and interaction strengths respectively.  $L$ is the number of sites and we consider the system at half-filling, i.e. the number of particles is $n=\sum_x n_x = L/2$.
This model is equivalent to the disordered XXZ Heisenberg spin-$1/2$ chain where the interaction $V$ corresponds to the anisotropy along and perpendicular to the $z$-axis.

\begin{figure}
\includegraphics[width=1.\columnwidth]{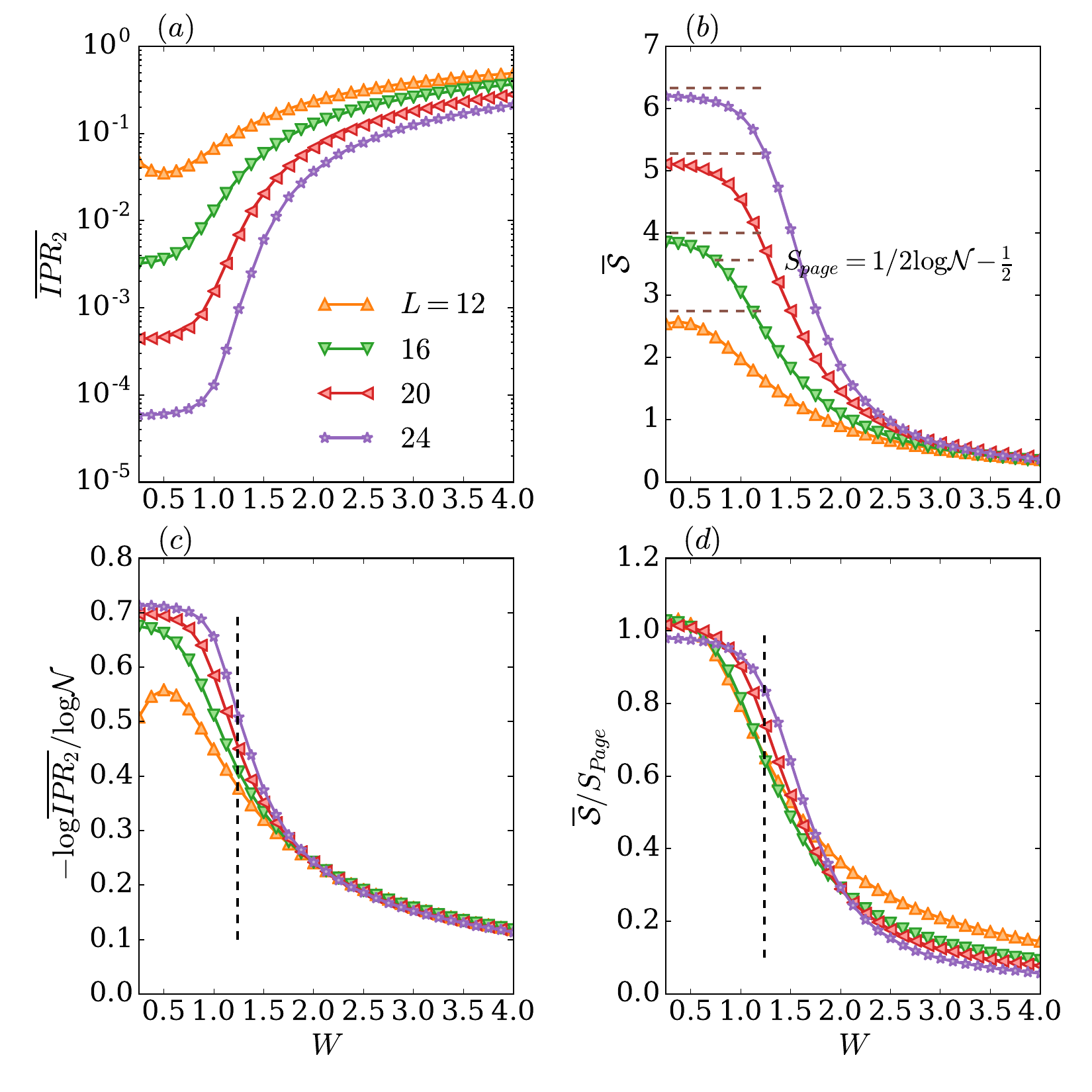}
 \caption{\textbf{IPR and EE for the many-body model with Hilbert space fragmentation as functions of disorder amplitude $W$.}
 (a)~IPR versus $W$ for different system sizes (shown in legend);
 (b)~EE $\overline{\mathcal{S}}$ versus $W$ for the same system sizes. Dashed horizontal lines show the Page value for the corresponding system size;
 (c)~Finite-size estimate $-\log \overline{IPR}/\log \mathcal{N}$ of the fractal dimension $D_2$ versus $W$;
 (d)~Finite-size flow of the $\overline{\mathcal{S}}/{\mathcal{S}}_{Page}$.
 In panels~(c) and~(d) vertical dashed lines correspond to the $D_2(L_{\max}) = 0.5$, where $L_{\max}=24$ is the maximal system size we calculated.
 }
 \label{fig:R1}
 \end{figure}
At finite interaction strength, $V=1$, this model is believed to have a many-body localization (MBL) transition
 at $W_c \approx 3.5$ ($W<W_c$ ergodic and $W>W_c$ localized). In our case we consider another limit of large interaction strengths, i.e. $V/t\to\infty$.  As shown in~\cite{detomasi2019dynamics}, $\hat H$ can be mapped to the following local Hamiltonian
\begin{equation}
\label{eq:H_inf}
 \hat H_{\infty} =   -t\sum_x  \hat P_{x} \big( \hat c_{x+1}^\dagger \hat c_{x} + \textit{h.c.} \big)\hat P_{x}  + W\sum_x \mu_x \hat n_x,
\end{equation}
with the dynamical constraints imposed by local projectors
\begin{equation}
 \hat P_{x} = 1 - (\hat n_{x+2} - \hat n_{x-1})^2, \qquad \hat P_{x}^2 = \hat P_{x} \ .
\end{equation}

In this limit, $\sum_x \hat n_x \hat n_{x+1}$ is a new conserved quantity, and the Hilbert/Fock space fragments in several disjoint blocks given by the value of $\sum_x \hat n_x \hat n_{x+1}$. We consider the largest block for which $\sum_x \hat n_x \hat n_{x+1} = L/4$ and the dimension of its Hilbert space is $\mathcal{N} = \binom{L/2}{L/4}^2\sim \frac{2^L}{L}$, thus up to polynomial corrections in $L$, it still spans the full Hilbert  space of $\hat H$, Eq.~\eqref{eq:full_H}. Nevertheless, even a further fragmentation takes place in this block, which disjoints it into exponential many blocks $\hat H_{\infty}$ (for details see Ref.~\cite{detomasi2019dynamics}). We focus our analysis on the largest sub-block, which has the dimension $n\binom{3n/2-2}{n/2} \sim \sqrt{L} 2^{D_{\text{Hilbert}} L}$, where $D_{\text{Hilbert}} = \frac{3}{4}\log_2{3} - \frac{1}{2}\approx 0.7$. Thus, the eigenstates of $\hat H_{\infty}$ are confined in an exponentially small fraction of the full Hilbert space and their fractal dimensions, $D_q$, should be always smaller than unity  $D_{\text{Hilbert}}<1$.
Using the Wigner-Jordan transformations, one can show that the $t{-}V$ is equivalent to the XXZ Heisenberg spin-$1/2$ chain, so we also expect that the above wave functions for $V\to\infty$ are multifractal as it has been shown for finite anisotropy in XXZ model in~\cite{luitz2019multifractality}.

Now, we show numerically that even though the fractal dimensions of the eigenstates of $\hat H_{\infty}$ are strictly smaller than one, the bipartite entanglement entropy $\mathcal{S}$ of these states can scale as the Page value $\sim L/2 \log{2}$.
Figures~\ref{fig:R1}~(a)-(b) show the disorder-averaged $\overline{IPR_2}$ and half-chain EE $\overline{\mathcal{S}}$ for eigenstates in the middle of the spectrum of $\hat H_{\infty}$ as a function of disorder strength $W$ (see also~\cite{detomasi2019dynamics} for more details). $\overline{IPR_2}$ decreases exponentially with $L$,  $\overline{IPR_2} \sim 2^{-D_2 L}$, $\mathcal{N}=2^L$. We can give an estimation of the fractal dimension $D_2$ considering its finite-size approximation $D_2(L) = -\log{\overline{IPR_2}}/\log{\mathcal{N}}$. Up to finite-size corrections, from the numerics we can deduce that $ 0.5 \le D_2\le D_1 \le D_{\text{Hilbert}}\approx 0.7$ for $W\le 1.25$ as shown in Fig.~\ref{fig:R1}~(c). At the same time, we can estimate the scaling of the EE $D_{ent}$ as $D_{ent}(L) = \overline{\mathcal{S}}/\mathcal{S}_{Page}$, see Fig.~\ref{fig:R1}~(d). As one can see, $D_{ent} \approx 1$ even though $D_1 \le 0.7$ at weak disorder. This gives numerical evidence of our statements for a quantum many-body system.

\end{document}